\documentclass[fleqn,usenatbib]{mnras}

\usepackage{newtxtext,newtxmath}
\usepackage[T1]{fontenc}

\DeclareRobustCommand{\VAN}[3]{#2}
\let\VANthebibliography\thebibliography
\def\thebibliography{\DeclareRobustCommand{\VAN}[3]{##3}\VANthebibliography}

\usepackage{graphicx}	% Including figure files
\usepackage{amsmath}	% Advanced maths commands
\title[MIGHTEE: multi-wavelength counterparts]{MIGHTEE: multi-wavelength counterparts in the COSMOS field}

\author[I. H. Whittam et al.]{I. H. Whittam$^{1,2}$\thanks{E-mail: imogen.whittam@physics.ox.ac.uk},
M. Prescott$^{2,3}$, C. L. Hale$^{4}$, M .J. Jarvis$^{1,2}$, I. Heywood$^{1,5,6}$,  \newauthor  Fangxia An$^{2,3}$, M. Glowacki$^{2,7,8}$, N. Maddox$^{9}$, L. Marchetti$^{10,11}$, L. K. Morabito$^{12,13}$, \newauthor N. J. Adams$^{3,14}$, R. A. A. Bowler$^{1,14}$, P. W. Hatfield$^{1}$, R. G. Varadaraj$^{1}$, J. Collier$^{8}$, \newauthor B. Frank$^{6}$, A.R. Taylor$^{2,7}$, M. G. Santos$^{2,6}$,  M. Vaccari$^{2,7,11}$, J. Afonso$^{15,16}$, Y. Ao$^{17,18}$, \newauthor J. Delhaize$^{10}$, K. Knowles$^{5,6}$,  S. Kolwa$^{7,19}$, S.M. Randriamampandry$^{20,21}$, \newauthor Z. Randriamanakoto$^{20,21}$, O. Smirnov$^{5,6}$, D. J. B. Smith$^{22}$ and S. V. White$^{5}$ 
\\
{\it Author affiliations are listed at the end of the paper.}
}

\date{Accepted XXX. Received YYY; in original form ZZZ}

\pubyear{2023}

\begin{document}
\label{firstpage}
\pagerange{\pageref{firstpage}--\pageref{lastpage}}
\maketitle

% Abstract of the paper
\begin{abstract}

In this paper we combine the Early Science radio continuum data from the MeerKAT International GHz Tiered Extragalactic Exploration (MIGHTEE) Survey, with optical and near-infrared data and release the cross-matched catalogues. The radio data used in this work covers $0.86$ deg$^2$ of the COSMOS field, reaches a thermal noise of $1.7$ $\muup$Jy/beam and contains $6102$ radio components. We visually inspect and cross-match the radio sample with optical and near-infrared data from the Hyper Suprime-Cam (HSC) and UltraVISTA surveys. This allows the properties of active galactic nuclei and star-forming populations of galaxies to be probed out to $z \approx 5$. Additionally, we use the likelihood ratio method to automatically cross-match the radio and optical catalogues and compare this to the visually cross-matched catalogue. We find that 94 per cent of our radio source catalogue can be matched with this method, with a reliability of $95$ per cent. We proceed to show that visual classification will still remain an essential process for the cross-matching of complex and extended radio sources. In the near future, the MIGHTEE survey will be expanded in area to cover a total of $\sim$20~deg$^2$; thus the combination of automated and visual identification will be critical. We compare redshift distribution of SFG and AGN to the SKADS and T-RECS simulations and find more AGN than predicted at $z \sim 1$.

\end{abstract}

% Select between one and six entries from the list of approved keywords.
% Don't make up new ones.

%%%%%%%%%%%%%%%%% BODY OF PAPER %%%%%%%%%%%%%%%%%%

\section{Introduction}\label{section:intro}

In order to truly understand the astrophysical processes that occur in our Universe, a multi-wavelength approach is necessary. This requires combining data from a number of different instruments operating across the full range of the electromagnetic spectrum. At the longest wavelengths, radio observations of extragalactic sources are invaluable; not only do they provide a dust-free view of star-forming galaxies (SFG), but they are also crucial for understanding Active Galactic Nuclei (AGN), which are powered by the supermassive black holes that reside in the centre of all massive galaxies, and are thought to play a key role in their evolution.

New radio facilities such as Meer-Karoo Array Telescope \citep[MeerKAT; ][]{Jonas2018, Mauch2020}, the Low-Frequency Array \citep[LOFAR; e.g.][]{VanHaarlem2013} and the Australian Square Kilometre Array Pathfinder \citep[ASKAP; e.g. ][]{Johnston2007, Hotan2021} are able to probe faint radio sources down to thermal noise levels of just a few $\upmu$Jy, which means we are no longer limited to observing the radio properties of only the brightest and most massive galaxies detected at optical wavelengths \citep[e.g.][]{Smolcic2017a,Heywood2020,Best2023}. Cross-matching radio and multi-wavelength data for these objects is necessary to build up a panchromatic view of the processes taking place in galaxies, which in turn allows us to determine their redshifts and other physical quantities such as luminosities and stellar masses.   

The MeerKAT International GHz Tiered Extragalactic Exploration (MIGHTEE) survey is one of the Large Survey Projects (LSPs) carried out with the MeerKAT telescope array. It will observe a number of well studied extragalactic fields, which have a wealth of multi-wavelength data available. These are the COSMOS, XMM-LSS, E-CDFS and ELAIS-S1 fields \citep{Jarvis2016}.
MeerKAT is being used to observe $20$ sq. deg. of sky, over a total of $\sim 1\,000$ hours of observation time, at L-Band radio frequencies between 856 -- 1712~MHz. The Early Science data release covering part of the COSMOS and XMM-LSS fields is described in \cite{Heywood2022}. 
As well as providing radio continuum images, MIGHTEE will also produce spectral line \citep{Maddox2021} and polarisation information (Sekhar et al. in prep.), allowing a range of science cases to be investigated. These include studying the evolution of star-forming galaxies and AGN, the role of AGN feedback in the quenching of star-formation, the evolution of neutral hydrogen in the Universe and measuring cosmic magnetic fields in large scale structures. 

Here we describe the process of cross-matching a subset of the Early Science MIGHTEE radio observations with multi-wavelength data in the COSMOS field. This paper is structured in the following way: in Section~\ref{Data} we describe the initial radio and multi-wavelength datasets that we cross-match. In Section~\ref{Cross-Matching} we lay out the method used to cross-match these two datasets using visual identification. Our visually inspected cross-matched catalogue is compared with those produced from the likelihood ratio method in Section~\ref{section:LR}. In Section~\ref{section:properties} we highlight the properties of the sample and discuss the reliability of the photometric redshifts of the radio sources. In Section~\ref{section:sims} we divide our sample into active galactic nuclei and star-forming populations and compare to predictions from simulations. We conclude in Section~\ref{Conclusions}.

Throughout this paper we assume the following cosmological constants: $H_{0} = 70$ kms$^{-1}$ Mpc$^{-1}$, $\Omega_{m} = 0.3$ and $\Omega_{\Lambda} = 0.7$. Unless stated differently all magnitudes are AB magnitudes.

\section{Data}\label{Data}

\subsection{Radio Data}\label{section:radio-data}

This work is based on the MIGHTEE Early Science continuum data in the COSMOS field. These data are described fully in \citet{Heywood2022} and summarised briefly below. The observations consist of a single pointing with the MeerKAT telescope centred on RA $10^\textrm{h}00^\textrm{m}28.6^\textrm{s}$, Dec $+02^\circ12\arcmin21\arcsec$. The full Early Science image covers 1.6~deg$^2$, but for this work we restrict ourselves to the central region with an area of 0.86~deg$^2$, where the radio data is deepest and approximately of uniform depth. The observations were taken between 2018 and 2020 with the L-band receiver (bandwidth \textrm{856 -- 1712 MHz}) and include 17.45 hours on source.

The MIGHTEE Early Science data contains two versions of the data processed with different \citet{Briggs1995Thesis} robust weighting values. The first `high-resolution' image is produced using a Briggs robust weighting of $-1.2$, which down-weights the short baselines in the core of the array. This results in a higher resolution of 5~arcsec, but comes at the expense of sensitivity, resulting in a $1\sigma$ thermal noise level of $6~\muup$Jy beam$^{-1}$. The second image uses a robust weighting of $0.0$, resulting in better sensitivity (thermal noise level of $1.7~\muup$Jy beam$^{-1}$) but a lower resolution of $8.6$~arcsec. Unlike the high resolution image, it should be noted that this lower resolution image is limited by classical confusion at the centre, meaning the actual measured noise is $4-5~\muup$Jy beam$^{-1}$. 

Source extraction on both images was conducted using the Python Blob Detection and Source Finder \citep[{\sc pybdsf},][]{Mohan2015}, as fully described in \citep{Heywood2022}. The primary catalogue we use in the cross-matching process here is the low resolution (Level 0) catalogue that contains $9\,915$ radio Gaussian components with peak brightnesses that exceed the local background noise by $5 \sigma_{\rm local}$. In this paper we crop the catalogue to remove sources away from the edge of the field and restrict the area to where the primary beam gain drops to $0.5$ resulting in a catalogue of $6338$ radio source components. We also remove 236 radio source components located within masked regions of the near-infrared image used for cross-matching (see Section~\ref{section:multi-data}). This results in a radio catalogue containing 6102 source components over an area of $\sim 0.86$~deg$^2$. A similar catalogue using the high resolution image contains 3116 radio source components over the same area. \cite{Heywood2022} also release a Level 1 catalogue based on the low-resolution image which has been visually inspected to remove artefacts and includes additional information. This work is based on the Level 0 catalogue, but we make use of the `resolved' flag in the Level 1 catalogue in Section~\ref{section:LR}. 

Complementary to our MIGHTEE observations are those of the VLA-COSMOS $3$ GHz Large Project \citep{Smolcic2017a}. Here the COSMOS field was observed in the S-band ($2$ - $4$ GHz) for a total of $384$ hours, in both the VLA's A and C-array configurations. The resulting image has resolution of $0.78\arcsec$ with a sensitivity of $2.3~\muup$Jy beam$^{-1}$. This is equivalent to a flux density of $\sim 4$\,$\muup$Jy\,beam$^{-1}$ at the mean effective frequency  of 1.34\,GHz for the lower resolution MIGHTEE Early Science data \citep{Heywood2022}. A total of 3949 of the 6102 components in the initial catalogue used in this work have a match within 8.6 arcsec in the VLA-COSMOS catalogue. This is discussed in more detail in \citet{Whittam2022}. 

\subsection{Multi-wavelength data}
\label{section:multi-data}

A wealth of multi-wavelength data for the COSMOS field has already been collated, and here we use the dataset fully described in \cite{Bowler2020, Adams2020,Adams2021}.  Covering $\sim 2$\,deg of the sky centred on the J2000 coordinates of RA = $10^\textrm{h}00^\textrm{m}28.6^\textrm{s}$ DEC = $+02^\circ12\arcmin21.0\arcsec$, this compilation includes $u^{*}$-band data from the Canada-France-Hawaii Telescope Legacy Survey \citep[CFHTLS,][]{Cuillandre2012}, $grizy$-band Hyper Suprime Cam (HSC) imaging \citep{Aihara2018}, near-infrared $YJHK_{s}$-band data from the UltraVISTA Survey \citep{McCracken2012}. Infrared data at $3.6$ and $4.5$ microns were obtained from the {\it Spitzer} Extended Deep Survey \citep{Ashby2013}. Source finding was conducted using {\sc SExtractor} \citep{Bertin1996}.  We adopt a flux-limited sample selected in the $K_{s}$ band with $K_{s} < 25$. We then carried out forced photometry in all other bands with the same fixed aperture and then adopt an aperture correction for determining the total flux from each object. Full details can be found in \citep{Adams2021}.

We use a compilation of spectroscopic redshifts from the following observing campaigns; VIMOS VLT Deep Survey \citep[VVDS,][]{Lefevre2013}, z-COSMOS \citep{Lily2009}, Sloan Digital Sky Survey \citep[SDSS DR12,][]{Alam2012}, 3D-HST \citep{Momcheva2016}, Primus \citep{coil2011}, and the Fiber-Multi Object Spectrograph \citep[FMOS,][]{Silverman2015}. Utilising the flag system provided by each survey, we ensure we only use spectroscopic redshifts which have a $>95$ per cent confidence of being correct. 

Photometric redshifts for the dataset were determined using a hierarchical Bayesian combination of two different techniques as conducted by \cite{Duncan2018}. The photometric redshifts were determined using a traditional template fitting technique carried out by the {\sc Le Phare} Spectral Energy Distribution (SED) fitting code \citep{Arnouts1999, Ilbert2006}, along with machine learning using the GPz algorithm \citep{Almosallam2016a,Almosallam2016b}. This method weights the combinations of photometric redshifts for both active galaxies and normal galaxies from the template fitting, and then combines this with the solutions determined from the more empirical machine learning approach with GPz. Full details and the catalogues can be found in \cite{Hatfield2020, Hatfield2022}. The photometric redshifts of the sources in our radio sample are discussed further in Section~\ref{section:properties}.

\section{Visual Cross-Matching}
\label{Cross-Matching}

The cross-matching of the radio and near-infrared datasets was carried out via visual inspection in a similar way to \cite{Prescott2018}. Overlays for each of the $6102$ radio components in the low resolution {\sc pybdsf} catalogue were produced using the Astronomical Plotting Library in Python \citep[APLpy,][]{Robitaille2012}. These overlays consist of radio contours produced from the MIGHTEE and $3$ GHz images overlaid on top of an UltraVISTA $K_{s}$-band image. The location of known sources from the near-infrared catalogue described in Section~\ref{section:multi-data} are also highlighted on top of the overlays. The cross-matching process is aided by the use of two different radio images with different resolution and sensitivity. The high resolution ($0.78$ arcsec) of the $3$ GHz VLA images allows a counterpart to be identified more easily, whereas the high sensitivity MeerKAT image reveals more diffuse radio sources. 
As in \cite{Prescott2018}, two sets of overlays are produced for each source to aid the visual classification. One overlay set has a size of $0.5\arcmin \times 0.5\arcmin$ whilst the other covers a larger area of $3\arcmin \times 3\arcmin$. The smaller overlay ensures we can assign the radio source with the correct counterpart for galaxies in crowded fields, and the larger overlays allows us to identify sources that are extended.   

In order to ensure we have a robust set of cross-matches, the radio sources were divided into batches of $100$ and inspected by three separate people from a team of $6$ classifiers. This was conducted using an improved version of the {\sc Xmatchit} code \citep{Prescott2018}, that now makes extensive use of Jupyter notebooks \citep{Kluyver2016}. When inspecting the overlays we classify the cross-matches as one of the following; 
\begin{itemize}
\item Single component - a single-component match, where the near-infrared counterpart to an isolated radio source is unambiguous.
\item Multiple-component - where multiple radio components are associated with a single near-infrared counterpart.
\item No visible optical counterpart - where the radio emission is not associated with a multi-component source and has no apparent near-infrared counterpart.
\item Confused source - where the resolution of the radio data is insufficient to identify an unambiguous counterpart. A subset of these sources are subsequently split into separate sources using the higher-resolution VLA 3~GHz data, as described below.
\end{itemize}

The output from each classifier was then compared to find mismatches. When mismatches occurred the overlays were re-inspected by a team of three experts and re-classified. Despite visual classification being a subjective and time consuming process, it is still necessary, as we show when comparing it to the likelihood ratio technique in Section~\ref{section:LR} and it is recognised as being more reliable than automated techniques \citep{Fan2015}. With visual classification, imaging and source detection errors can be noticed easily, and rare and interesting objects such as giant radio galaxies \citep[e.g.][]{Delhaize2021} can be identified.  

Peak and integrated radio flux densities for the single component sources in the cross-matched catalogue are directly taken from the low-resolution Level 0 MIGHTEE {\sc pybdsf} catalogue. Integrated flux densities for multi-component sources are the sum of integrated flux densities of the individual components, and the peak fluxes for multi-component sources are taken as the peak flux of the component with the highest peak flux.

For confused radio sources, if the radio source clearly consists of two or more radio sources that are separate sources in the VLA $3$~GHz catalogue and each have a separate host galaxy, we split the MIGHTEE radio source into two or more sources with separate near-infrared counterparts. We estimate the 1.3-GHz peak and integrated flux densities of these split sources by dividing the flux of the original MIGHTEE source into two (or more) according to the ratio of the fluxes of the VLA 3 GHz sources as follows

\begin{equation}
   S_{\textrm{1.3 }i} = S_\textrm{1.3 orig} \frac{S_{\textrm{3 } i}}{\sum_{i=1}^{N} {S_{\textrm{3 } i}} }
\end{equation}

\noindent where $S_\textrm{1.3 i}$ is the estimated 1.3~GHz flux density of the $i$-th split source, $S_\textrm{1.3 orig}$ is the original MIGHTEE flux density of the confused source, $S_\textrm{3 i}$ is the VLA 3~GHz flux density of the $i$-th split source and $N$ is the total number of sources the source is being split into. We note that this assumes that all of the confused components have a similar spectral index between the 3\,GHz data in VLA-COSMOS and the 1.3\,GHz MIGHTEE data. As these are generally faint radio sources, and are thus likely star-forming galaxies \citep[see][]{Whittam2022,Smolcic2017b}, this assumption would not produce a large systematic offset in flux density, as star-forming galaxies tend of have similar spectral indices of $\alpha \sim 0.7$. We note, in particular, that the peak fluxes scaled in this way should be used with caution. Confused MIGHTEE sources which cannot be clearly separated in this way are flagged as being too confused. A full breakdown of all the possible cross-matching outcomes and their flags can be seen in Table~\ref{tab:xmatch-classifications}. 

\begin{table*}
\centering
\caption{Breakdown of the classifications from the visual inspection of the MIGHTEE radio sources.}
\label{tab:xmatch-classifications}
\begin{tabular}{lcc}
\hline
Cross-matching classification  & Number of radio sources & Number of components \\
\hline
Single component radio sources with a counterpart identified  & $4881$ & $4881$ \\
Multi-component radio sources with a counterpart identified & $62$ & $264$ \\
Matched sources with split fluxes & $280$ & $137$\\
\hline
Radio sources with no counterpart visible & $144$ & $144$ \\
Radio sources too confused to identify a counterpart & $664$ & $664$  \\
Radio artefacts & $12$ & $12$ \\
\hline
Total & $6043$ & $6102$  \\\hline
\\
\end{tabular}
\end{table*} 

\begin{figure*}
\includegraphics[width =0.48\textwidth]{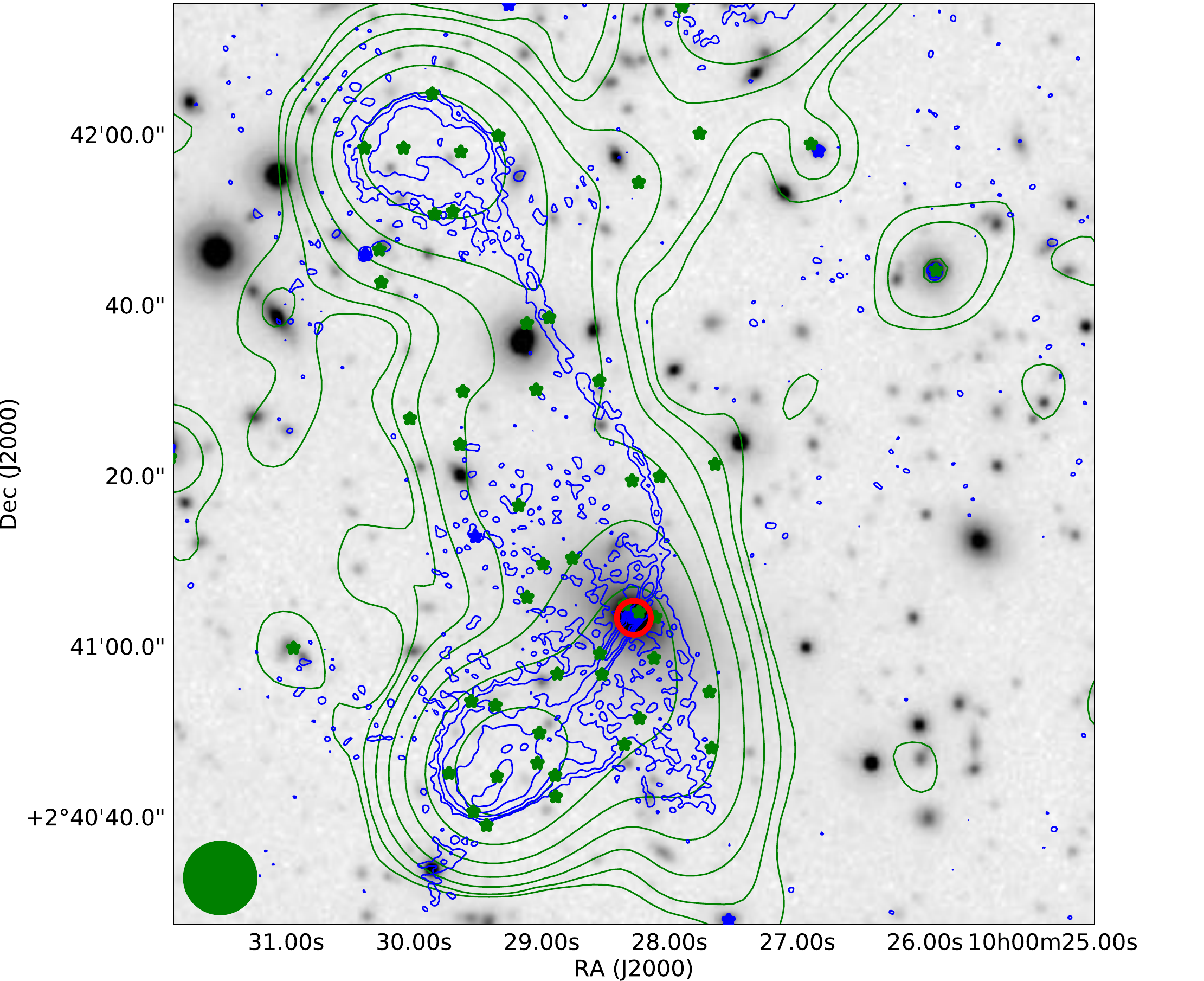}
\includegraphics[width =0.48\textwidth]{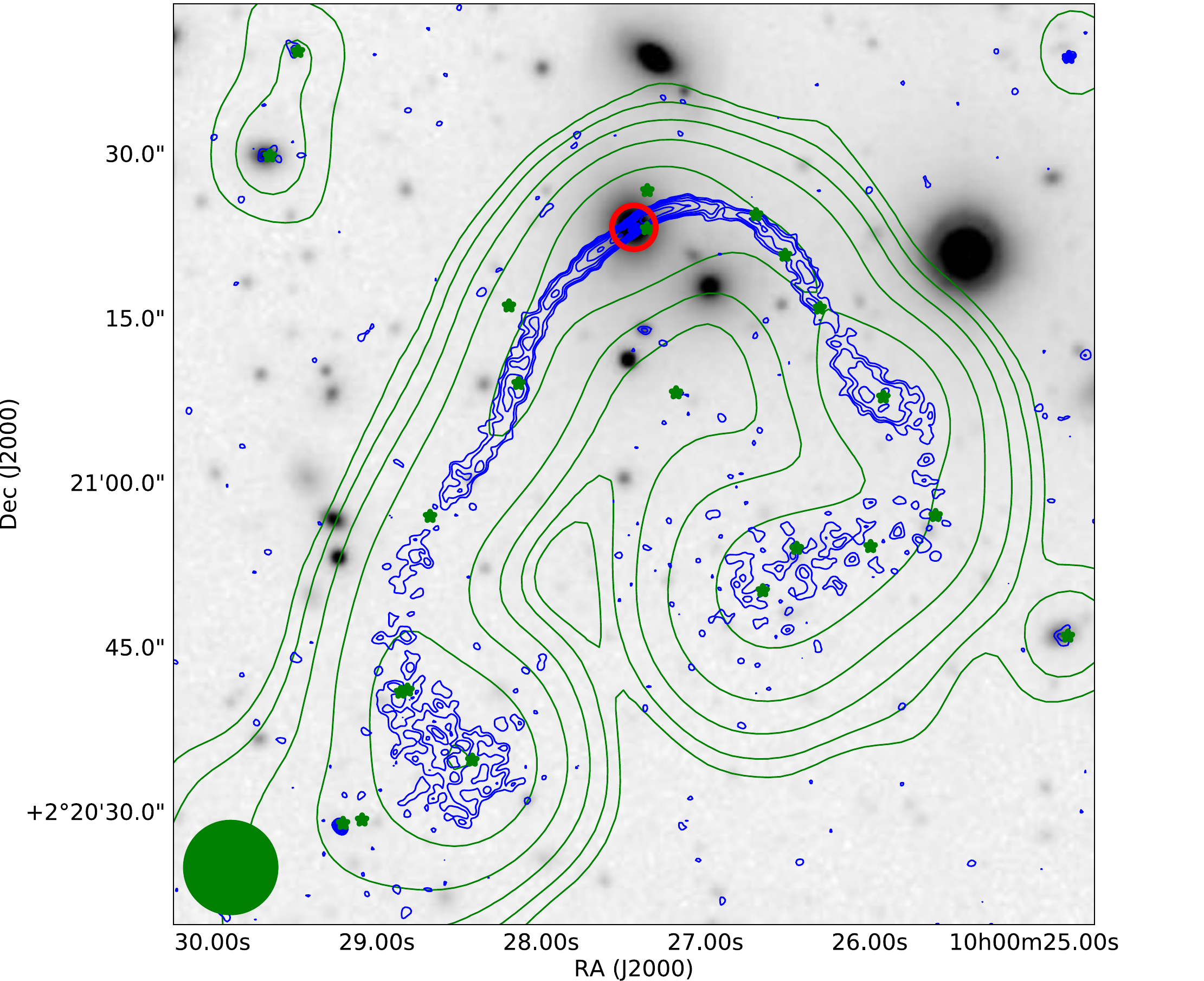}
\includegraphics[width =0.48\textwidth]{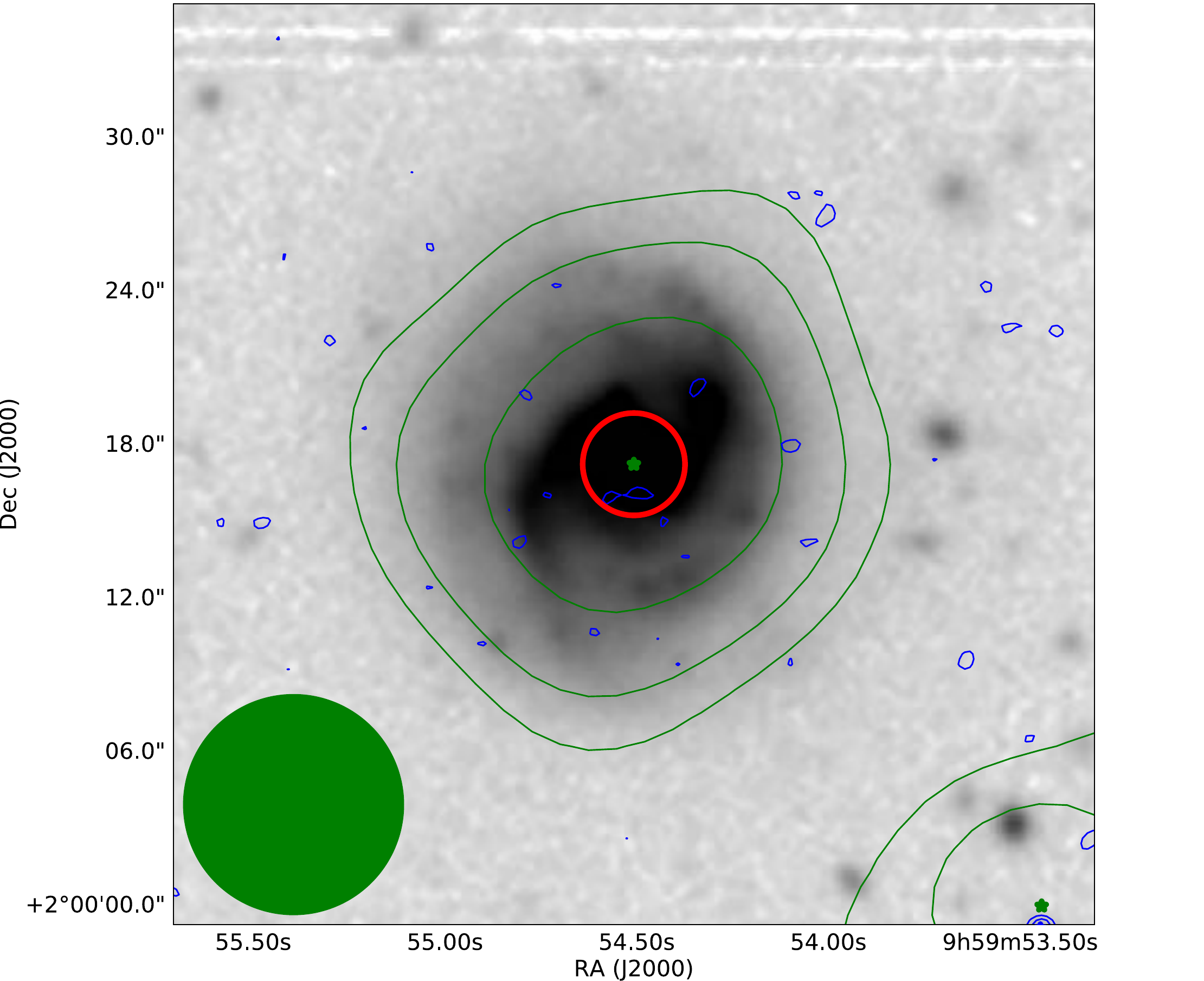}
\includegraphics[width =0.48\textwidth]{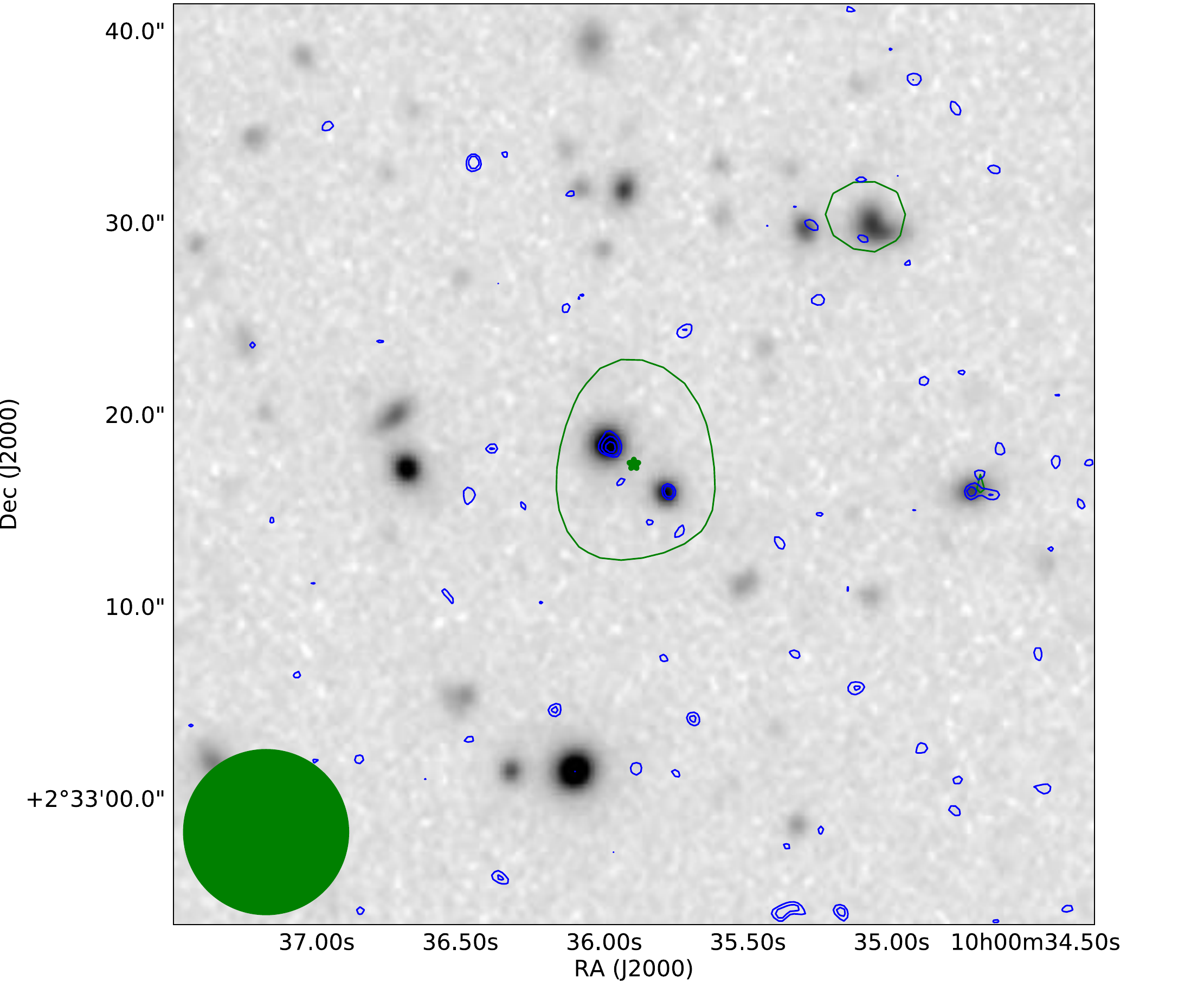}
\caption{Examples of overlays examined in the cross-matching process. Radio contours from MIGHTEE Early Science (\citealt{Heywood2022}, green) and the VLA-COSMOS $3$ GHz Large Project (\citealt{Smolcic2017a}, blue) are overlaid on an UltraVISTA K${_s}$-band background grey scale image \citep{McCracken2012}. The contour levels here represent $7$ levels evenly spaced in log space between $1.5$ times the local rms noise and half the maximum pixel value in the image. The green stars indicate radio components in the {\sc pybdsf} catalogue. The red circles indicate the host galaxies of these radio sources. The upper two panels show two large extended, multi-component AGN. The bottom left panel a single component star-forming galaxy, and the bottom right panel shows a radio source that is confused in MIGHTEE. The size of the MeerKAT radio beam is highlighted by the solid green circle.}
\label{EXAMPLES}
\end{figure*}

Examples of the different classifications from the cross-matching process can be seen in Fig.~\ref{EXAMPLES}. The green and blue contours show the MIGHTEE and VLA-COSMOS 3 GHz Survey imaging data respectively, overlaid on a grey scale UltraVISTA $K_{s}$-band image. The upper left and right panels display two large extended AGN residing in host galaxies at redshifts of $z=0.349 $ and $z=0.219$, that are made up of multiple radio components ($N_\textrm{comp}=47$ and $N_\textrm{comp}=17$). The bottom left panel displays a nearby ($z=0.078$) star-forming galaxy, comprising of a single radio component. The bottom right panel highlights a confused source, where two objects are contributing flux to a single MIGHTEE component radio source.  This source has been split into two separate sources in the resulting visual cross-matched catalogue, with 1.3-GHz flux densities estimated from the 3-GHz flux densities as described above. 

A total of $5\,282$ of the initial {\sc pybdsf} catalogue of radio components could be visually matched to $5\,223$ $K_{s}$-band counterparts. Note that there is not a direct mapping between sources in the input and cross-matched catalogues, as components which form part of multi-component sources have been combined and some blended sources have been split. The percentage of the initial radio components we can cross-match is therefore $87$ per cent ($5\,282$ out of $6\,102$).  This appears to be an improvement over previous studies, for example \cite{Prescott2018} found that only $57$ per cent of their initial radio catalogue from the VLA Stripe 82 Snapshot Survey \citep{Heywood2016} could be cross-matched to an optical source and \cite{Williams2019} found $73$ per cent of their radio sources from the LOFAR Two-metre Sky Survey (LoTSS) have optical/IR identifications from Pan-STARRS and/or WISE. However, due to the shallower radio and multi-wavelength datasets used by these studies, the samples are not directly comparable, as the ability to identify counterparts to radio sources is influenced by the depth of both the radio and the multi-wavelength imaging. A more useful comparison is to the recent LOFAR Deep Fields work by \citep{Kondapally2021}. They cross-match the LOFAR deep field data to a wealth of multi-wavelength imaging data and achieve a successful identification for 97 per cent of the radio sources over the three deep fields using a combination of visual identification and automated cross-matching, and we return to this in Section~\ref{section:LR}.

\begin{figure}
\includegraphics[width=\columnwidth]{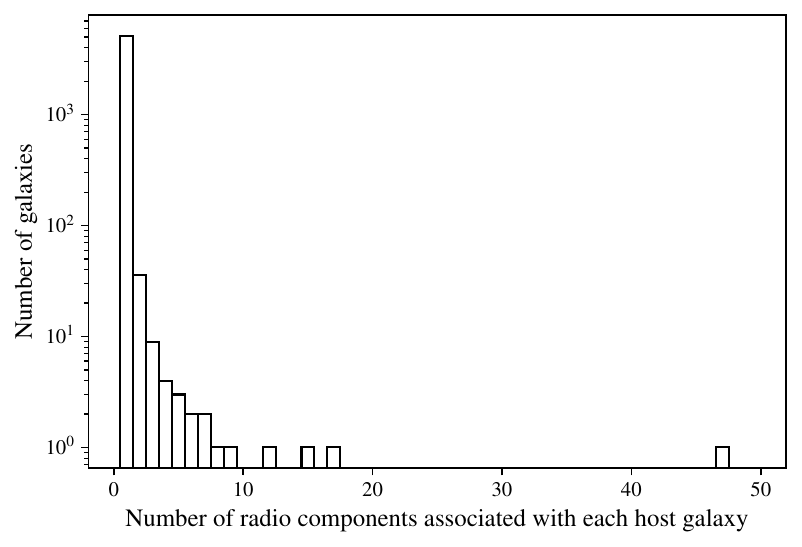}
\caption{Histogram of the number of MIGHTEE radio components belonging to a single near-infrared object. The vast majority of cross-matched counterparts in our survey comprise of a single radio component.}
\label{FIGNCOMP}
\end{figure}

\begin{figure}
\includegraphics[width=\columnwidth]{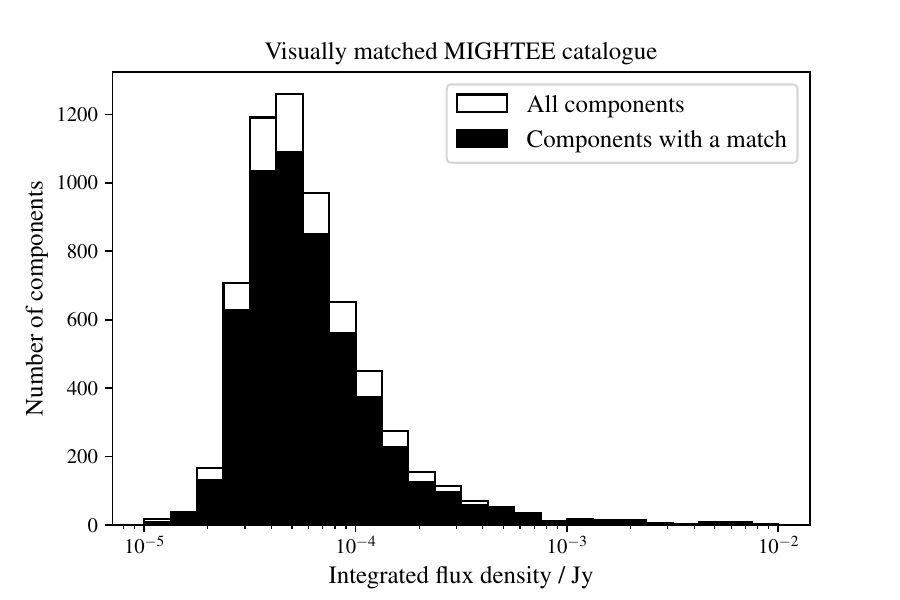}
\includegraphics[width=\columnwidth]{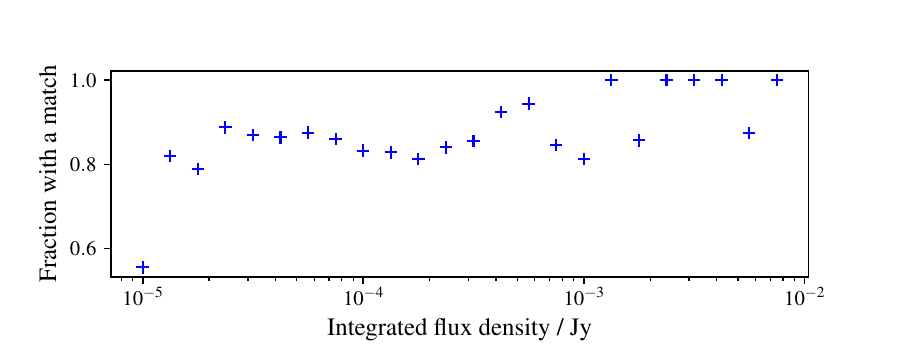}
\caption{The distribution of total fluxes of all components in the MIGHTEE catalogue (white), with those with a counterpart in the visually matched catalogue shown in black. The bottom panel shows the fraction of matched components in each flux density bin.}
\label{fig:match-flux}
\end{figure}

The numbers of radio components that have been assigned to a single optical counterpart can be seen in Fig.~\ref{FIGNCOMP}. This shows the vast majority of objects ($99$ per cent) are comprised of a single radio component, 
while a small number are very extended, with 10 sources consisting of $>5$ components. These extended, multi-component sources are particularly challenging to match automatically and demonstrate the benefit of identifying counterparts by eye. This not only allows us to identify an appropriate host galaxy for the radio source, but also enables us to combine all detected components into one source, meaning we can produce a reliable estimate of the total source flux. The fraction of cross-matched sources in each radio flux density bin is shown in Fig.~\ref{fig:match-flux}. This shows that although there is not a strong dependence on our ability to visually cross-match sources as a function of their flux-density, we are more successful at identifying counterparts for the brightest sources. When we consider only sources with $S_{1.4~\textrm{GHz}} > 0.4$~mJy, the match fraction rises to 97 per cent.

\begin{figure}
\includegraphics[width=\columnwidth]{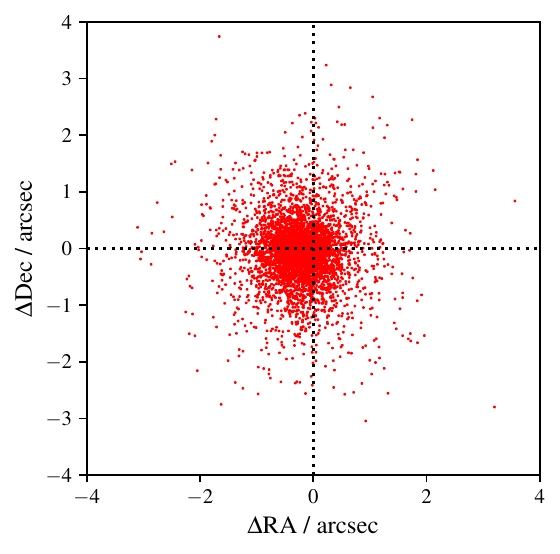}
\caption{The positional offsets between the radio and $K_s$-band coordinates for each of the single component radio sources in the cross-matched catalogue.}
\label{OFFSETS}
\end{figure}

The positional offsets between the radio and $K_s$-band coordinates of our single radio component cross-matches can be seen in Fig.~\ref{OFFSETS}. The mean offset between the radio and $K_s$-band cross-matches is $0.24$~arcsec in RA and $0.40$~arcsec in Dec. As these offsets are significantly less than the resolution of the radio data, we do not correct for them in the cross-matching analysis.

\begin{figure}
\includegraphics[width=\columnwidth]{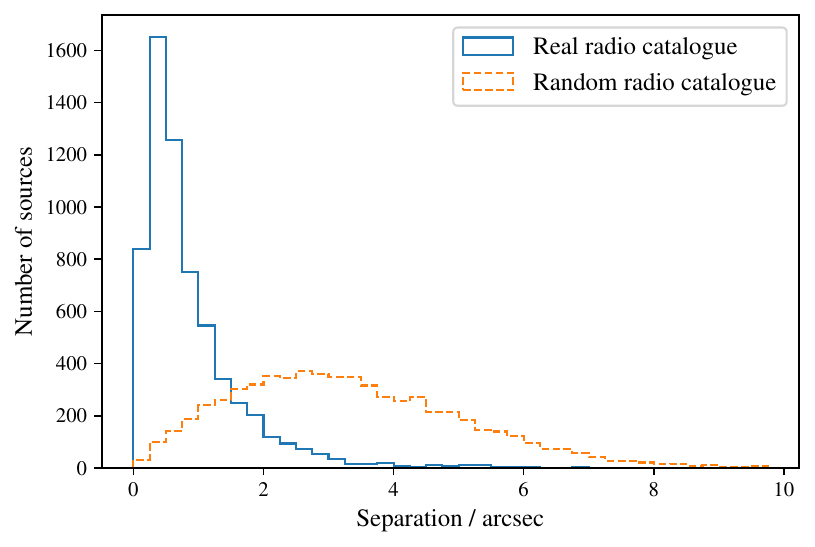}
\caption{Separation between each radio source and the nearest object in the $K_s$-band near-infrared catalogue for the real radio catalogue and a random set of positions with the same source density as the radio catalogue.
}
\label{fig:match-separation}
\end{figure}

In order to test the robustness of our visual cross-matching process, we employ a similar method to \cite{Prescott2016} and \cite{Prescott2018}. We measure the separation between each component in our input radio catalogue and the nearest object in the near-infrared catalogue. We then repeat this process with a catalogue of random radio positions, generated to have the same source density as the real radio catalogue. The resulting distribution of separations between the real and random radio sources and the nearest near-infrared source is shown in Fig.~\ref{fig:match-separation}. If we only consider cases where the separation between the radio source and the match in the near-infrared catalogue is less than 1~arcsec, there are 4501 matches identified to the real radio catalogue and 456 to the random catalogue, giving a reliability of 90 percent and a completeness of 71 per cent. Setting the separation limit at 2~arcsec raises the completeness to 92 per cent, but this is at the expense of reliability which drops to 73 per cent. Thus, use of the visually cross-matched catalogue should be tailored according to the science that is being carried out, and choosing the appropriate balance between reliability and completeness. Our final catalogue contains $4\,881$ matched sources comprising of a single radio component and $62$ matched multi-component radio sources. There are a further 280 split matches, giving a total of 5223 sources in the visually matched catalogue.

A description of the columns of the visually cross-matched (Level 2) catalogue, released with this work, can be seen in Appendix \ref{appendixa}. A catalogue of source classifications based on these visually cross-matched sources and their multi-wavelength data (the Level 3 catalogue) was released with \citet{Whittam2022}.

\section{The Likelihood Ratio}
\label{section:LR}

In this section we show how our visually inspected cross-matched catalogue compares to the result of an automated method and highlight the advantages and disadvantages of both methods.

The likelihood ratio (LR) describes the ratio of the probability that a given radio source is related to a particular optical/infrared counterpart to the probability that it is unrelated \citep{Sutherland1992} given by:  

\begin{equation}
{\rm LR} = \frac{q(m)f(r)}{n(m)},
\end{equation}

\noindent where $q(m)$ is the expected distribution of the true counterparts as a function of optical/infrared magnitude. $f(r)$ is the radial probability distribution function of the offsets between the radio and optical/infrared positions, and $n(m)$ is the magnitude distribution of the entire catalogue of optical/infrared detected objects.

This has been used by a number of studies to identify the multi-wavelength counterparts to radio catalogues, and can be very effective for single, isolated sources \citep{Smith2011,McAlpine2012,Kondapally2021}. Following the method described in \cite{McAlpine2012} (which contains a detailed description of how each of the terms in the equation above are calculated), we use the likelihood ratio to identify the host galaxies of the radio sources in both our high and low resolution catalogues, and use our visually cross-matched catalogue to evaluate the success of this method for the MIGHTEE COSMOS field. The ultimate aim is to determine whether the likelihood ratio can be used to match a sub-sample of the MIGHTEE sources automatically, thereby reducing the total number of sources which need to be matched by eye for the rest of the survey. This will be important given the much larger area which is yet to be cross-matched (this paper concerns less than 1~deg$^2$ out of a total $\sim 20$~deg$^2$).\footnote{The LR code used in this work can be found at: \url{https://github.com/lmorabit/likelihood_ratio}.}

\begin{figure}
    \centering
    \includegraphics[width=\columnwidth]{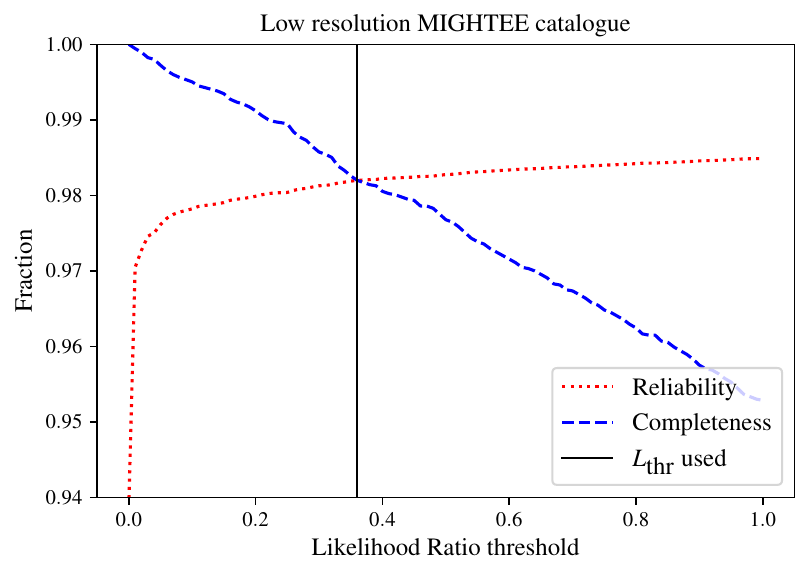}
    \includegraphics[width=\columnwidth]{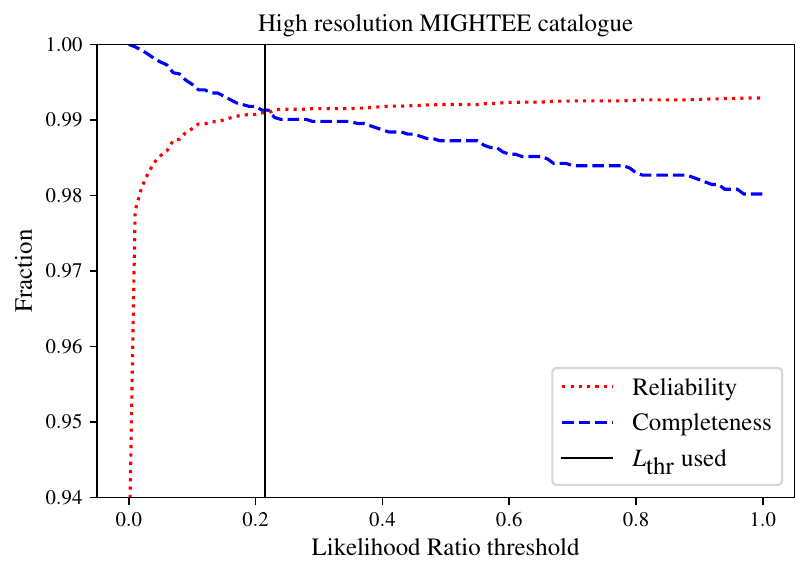}
    \caption{The completeness and reliability as a function of Likelihood Ratio threshold ($L_\textrm{thr}$) for the low and high resolution MIGHTEE catalogues (top and bottom panels respectively). The threshold used in this work is shown by the solid black line.}
    \label{fig:LRthresh}
\end{figure}

The input radio catalogues for the likelihood ratio method are the Level-0 {\sc pybdsf} source catalogues produced from both the high and low resolution MIGHTEE Early Science radio images, cut to the same 0.86 deg$^2$ area as used for the visual cross-matching (see Section~\ref{section:radio-data}). Although the visual cross-matching described in the previous section is based on only the low-resolution catalogue, here we employ the LR method to both the low and high resolution catalogues. This is because a cross-matched high-resolution catalogue has useful science applications, and because it allows us to inform our cross-matching strategy for different resolution images for the full MIGHTEE survey.  We search for counterparts in an UltraVISTA $K_{s}$-band selected catalogue with $K_{s}<25$. For those sources detected in the $K$, $G$, $i$ and $J$ bands using magnitude limits of 25.0, 27.4, 26.9, 26.6 respectively we find stars using the stellar locus defined in \cite{Jarvis2013}. Our final IR catalogue contains all objects in the initial IR catalogue with stars removed and with $K_s < 25$.
 For each radio source we select the object with the highest LR, and retain this match provided the LR value is above our defined threshold, $L_\textrm{thr}$. To determine the most appropriate LR threshold to use, we calculate the completeness and reliability for a given $L_\textrm{thr}$ in a similar way to \citet{Williams2019}

\begin{equation}
    C(L_\textrm{thr}) = 1 - \frac{1}{Q_0 N_\textrm{radio}} \sum_{\textrm{LR}_i < L_\textrm{thr}} \frac{Q_0 \textrm{LR}_i}{Q_0 \textrm{LR}_i + (1 - Q_0)},
\end{equation}

\begin{equation}
    R(L_\textrm{thr}) = 1 - \frac{1}{Q_0 N_\textrm{radio}} \sum_{\textrm{LR}_i \geq L_\textrm{thr}} \frac{1 - Q_0}{Q_0 \textrm{LR}_i + (1 - Q_0)},
\end{equation}

\noindent where $C(L_\textrm{thr})$ is the completeness for a given $L_\textrm{thr}$ (i.e.\ the fraction of real matches which are accepted) and $R(L_\textrm{thr})$ is the reliability (i.e.\ the fraction of accepted matches which are correct). $Q_0$ represents the fraction of radio sources which have a counterpart, $Q_0 = N_\textrm{matched} / N_\textrm{radio}$ which we calculate following the method outlined in \cite{Fleuren2012}. Following \citet{Williams2019} we set our LR threshold to the point where the $C(L_\textrm{thr})$ and $R(L_\textrm{thr})$ curves intersect. This gives us LR threshold values of 0.22 and 0.36 for the high and low resolution MIGHTEE catalogues respectively. The completeness and reliability curves as a function of $L_\textrm{thr}$ are shown in Fig.~\ref{fig:LRthresh}.  

\begin{table*}
\centering
\caption{Summary of the performance of the likelihood ratio to identify counterparts for radio sources in the high and low resolution MIGHTEE images.}
\label{tab:LRcomparison}
\begin{tabular}{lcc}
\hline
Radio Catalogue & Low resolution & High resolution  \\
\hline
Number of radio components in area covered by visual cross-match and LR  & $6102$ & $3116^1$ \\
Number with a good LR match (LR > $L_\textrm{thr}$) & $5747$ & $2916$ \\\hline
\% of input radio components with a good LR match & 94.2 & 93.6 \\\hline
Number of input components with a good visual match & $5408^2$ & $2759$ \\
\% of input radio components with a good visual match & 88.6 & 88.5 \\\hline
Number with both a good LR match and a good visual match & $4929$ & $2494$ \\
Number where matches from the two methods agree & $4657$ & $2381$ \\\hline
\% of components with good matches from both methods where the matches agree & $4657/4929 = 94.5$ & $2381/2494 = 95.5$ \\
\% of matched LR components where the matches from the two methods agree & $4657/5747 = 81.0$ & $2381/2916 = 81.7$ \\
\% of input components where the matches from the two methods agree & $4657/6102 = 76.3$ & $2381/3116 = 76.4$\\\hline\hline
Number of unresolved components in area covered by visual cross-match and LR & $5572$ & 3029\\
Number of unresolved components with a good LR match (LR > $L_\textrm{thr}$) & $5428$ & 2884\\\hline
\% of unresolved components with a good LR match & $97.4$ & 95.2 \\\hline
Number of unresolved components with both a good LR match and a good visual match & 4725 & 2469\\
Number where matches from the two methods agree & 4483 & 2365\\
\hline
Notes: & & \\
\multicolumn{3}{l}{$^1$ The high-resolution image is less sensitive so the resulting catalogue contains fewer sources than the low-resolution catalogue.}\\
\multicolumn{3}{l}{$^2$ In the final visual cross-matched catalogue components of multi-component sources have been combined and some blended sources have been split,}\\
\multicolumn{3}{l}{resulting in 5223 sources in the final catalogue.}\\
\end{tabular}
\end{table*} 

\subsection{The likelihood ratio for all sources}

\begin{figure}
\includegraphics[width=\columnwidth]{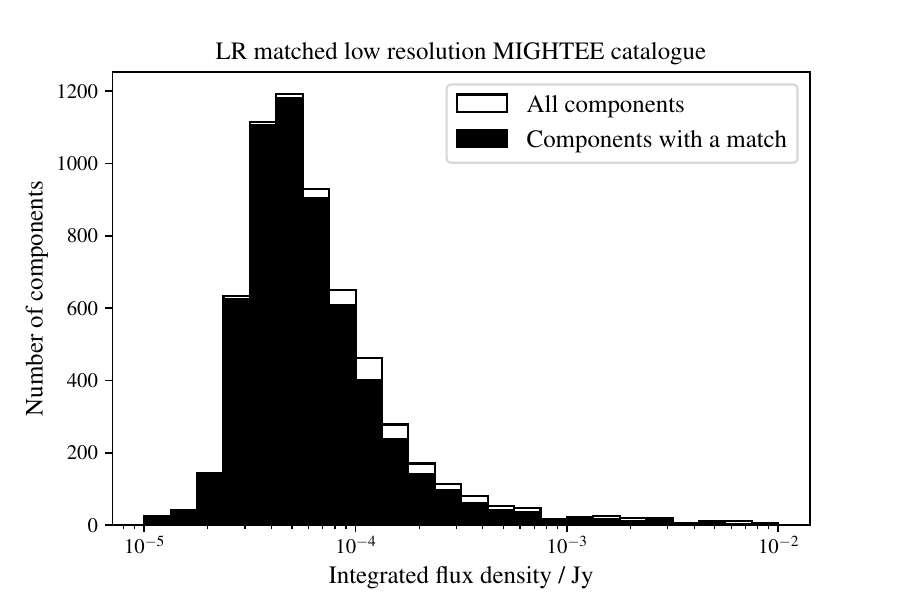}
\includegraphics[width=\columnwidth]{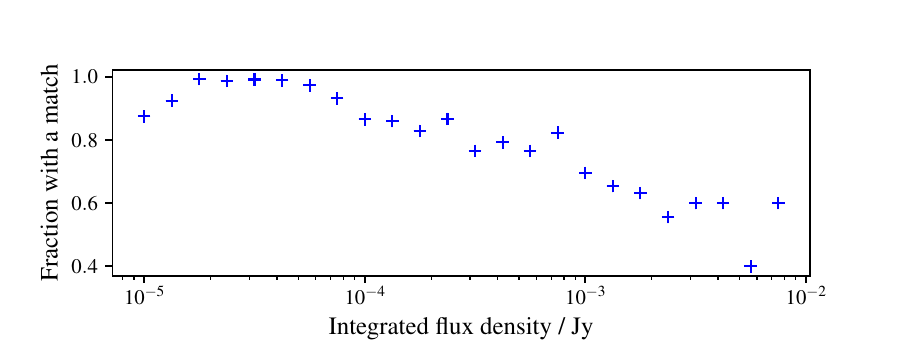}
\caption{The distribution of total fluxes of all components in the low resolution MIGHTEE catalogue (white), with those with a counterpart in the LR matched catalogue shown in black. The bottom panel shows the fraction of matched in each flux density bin.}
\label{fig:match-flux_LRlow}
\end{figure}

\begin{figure}
\includegraphics[width=\columnwidth]{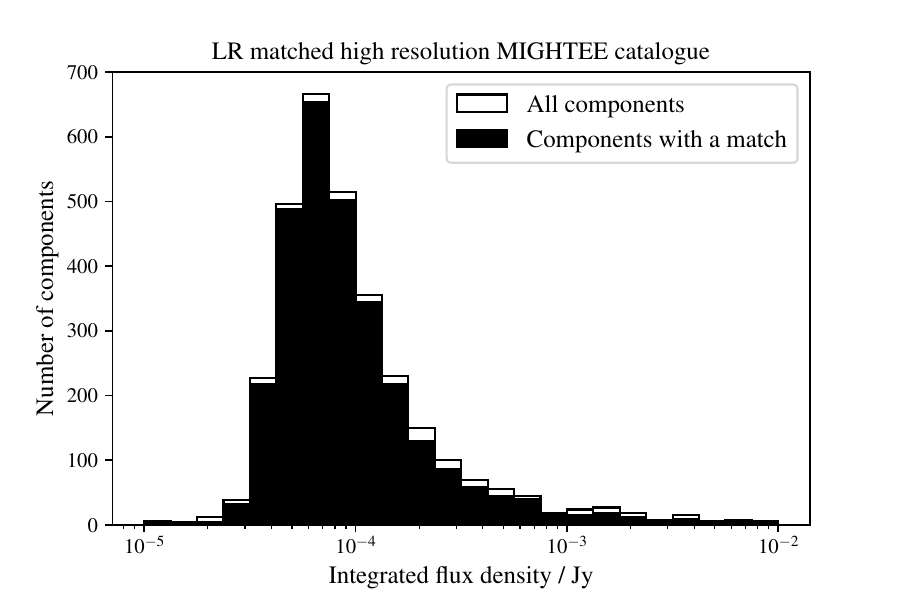}
\includegraphics[width=\columnwidth]{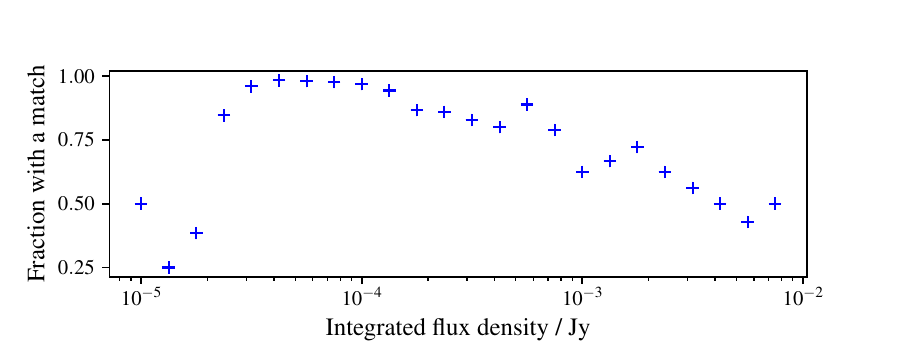}
\caption{The distribution of total fluxes of all components in the high resolution MIGHTEE catalogue (white), with those with a counterpart in the LR matched catalogue shown in black. The bottom panel shows the fraction of matched in each flux density bin.}
\label{fig:match-flux_LRhigh}
\end{figure}

Table~\ref{tab:LRcomparison} shows the performance of the likelihood ratio method on our radio source catalogues. With the likelihood ratio we are able to identify counterparts for 93.6 and 94.2 per cent of the initial high and low resolution radio component catalogues respectively. Figs.~\ref{fig:match-flux_LRlow} and \ref{fig:match-flux_LRhigh} show the flux density distribution of the sources in the MIGHTEE catalogue we are able to match using the LR method, and the fraction of matches in each flux bin. This demonstrates that the LR method is less successful at higher flux densities, due to the larger fraction of complex sources as discussed above. This is in contrast to the match fraction for the visually cross-matched catalogue shown in Fig.~\ref{fig:match-flux}, which increases at larger flux densities. For sources with $S_{1.4~\textrm{GHz}} > 100~\muup$Jy, by matching visually we are able to identify a counterpart for 93 per cent of sources, while the LR method is only able to cross-match 61 per cent of the same sample. This highlights the benefit of combining the two methods; by using the LR we are able to automatically match a large number of the fainter sources, but it is still necessary to match the more complex sources, which tend to have larger flux densities, by eye.

For the sources which also have a good match in the visually cross-matched catalogue, the two methods identify the same counterpart for 95.5 and 94.3 per cent of sources in the high and low resolution catalogues respectively. Note that when an input radio source has been split into two or more sources with separate near-infrared counterparts when visual cross-matching (see Section~\ref{Cross-Matching}) this is automatically counted as a disagreement with the LR method, as both counterparts are not identified by the LR method. This highlights one important aspect of where the LR method can be misleading, as it will produce a high LR counterpart to a "single source" and be seen as successful, whereas the source itself is confused and has two optical/NIR counterparts. Such sources are readily identified in the visual classification. On the other hand, if higher resolution radio data was available then the radio source itself would have been split into separate components and the LR could have been successful in assigning two optical/NIR counterparts. 

\begin{figure}
\includegraphics[width=\columnwidth]{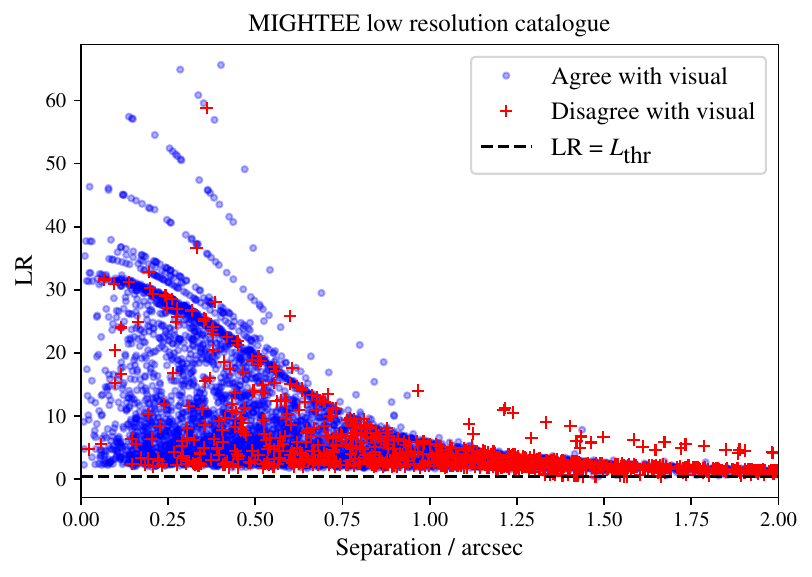}
\includegraphics[width=\columnwidth]{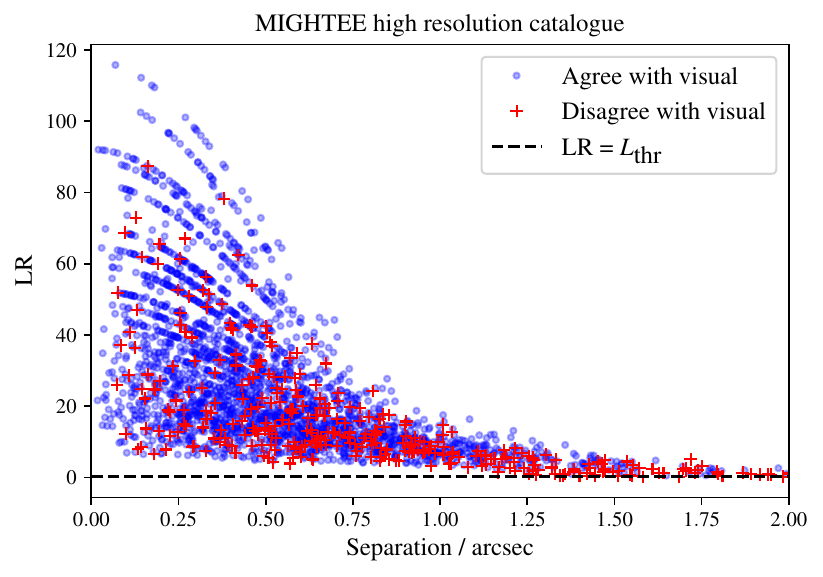}
\caption{The likelihood ratio as a function of separation between the radio source and multi-wavelength counterpart. The upper panel shows the likelihood ratio matches from the low resolution radio catalogue whilst the lower panel shows the high resolution radio catalogue. Sources where the counterpart identified using the LR method agree with that identified visually can be seen as blue circles, and those where they do not can be seen as red crosses. The black dashed line shows where $\textrm{LR} = L_\textrm{thr}$.}
\label{sepLR}
\end{figure}

However, this shows that the likelihood ratio method can be used to successfully identify counterparts for a large fraction of the MIGHTEE radio sources, and that the performance on the high and low resolution MIGHTEE catalogues is similar. For the sources with a good LR match, the two methods identify the same counterpart for 81.0 and 81.7 per cent of sources in the low and high resolution catalogues respectively. The likelihood ratio as a function of separation between the radio and near-infrared source positions can be seen in Fig.~\ref{sepLR}. The upper panel displays the likelihood ratio for the low resolution catalogue and the lower panel displays the same for the high resolution catalogue. The number of sources where the two methods disagree is higher when the separation between the radio and near-infrared positions are larger, and when the LR is lower, as expected.

We release the full likelihood ratio matched catalogues with this work and details can be found in Appendix~\ref{appendixa}.

\subsection{The likelihood ratio for unresolved sources}

We expect the likelihood ratio method to be more successful at identifying the correct counterpart for single, isolated sources than for extended sources, therefore we investigate whether excluding extended sources can increase the reliability of this method. As described in Section 3.3.3 of \citet{Heywood2022}, sources in the MIGHTEE Early Science catalogue are flagged as resolved if their deconvolved major axis size ($\phi_M$) exceeds the full-width half maximum of the restoring beam ($\theta_\textrm{beam}$) by 

\begin{equation}
    \phi_M - \theta_\textrm{beam} \geq 2\sigma_{\phi_M},
\end{equation}

\noindent where $\sigma_{\phi_M}$ is the uncertainty on the deconvolved major axis. There are 5572 sources in the low resolution catalogue which are not flagged as resolved, and 5429 (97.4 per cent) of these have a match identified by the likelihood ratio method described above. This demonstrates that the likelihood ratio method is able to cross-match a higher fraction of compact sources, as expected. 4725 of these sources also have a match identified in our visual classification catalogue, and for 4483 of these sources the counterparts identified by the two methods are the same object (this is 82.5 per cent of the 5428 unresolved components with a good LR match). The agreement of these matches with the visual classifications is therefore very similar to when we consider the full sample.

Despite the likelihood ratio on its own not being sufficient to identify multi-wavelength counterparts for each and every one of the MIGHTEE sources, it can be used successfully to produce a sub-sample of matched MIGHTEE sources and therefore dramatically reduce the total number of sources which need to be cross-matched by eye.  Obviously in any method there will be mismatches between the radio and the optical identifications due to the plethora of different structures seen in the radio, e.g. jets, lobes and hotspots from active galactic nuclei, and automating such cross-matching is extremely difficult. Thus, the need to use a combination of LR and visual cross-matching will remain and the adopted threshold to "eyeball" sources will necessarily change depending on the science which is being carried out, e.g. a balance between completeness and reliability.
We will use this analysis to inform the cross-matching strategy for the remaining MIGHTEE fields. For the cases where the visual cross-matches and the LR matches disagree, we would require additional information to be able to associate these sources, e.g. higher-resolution radio data or spectroscopy.

\section{Redshifts for the cross-matched sample}
\label{section:properties}

\begin{figure}
\includegraphics[width=\columnwidth]{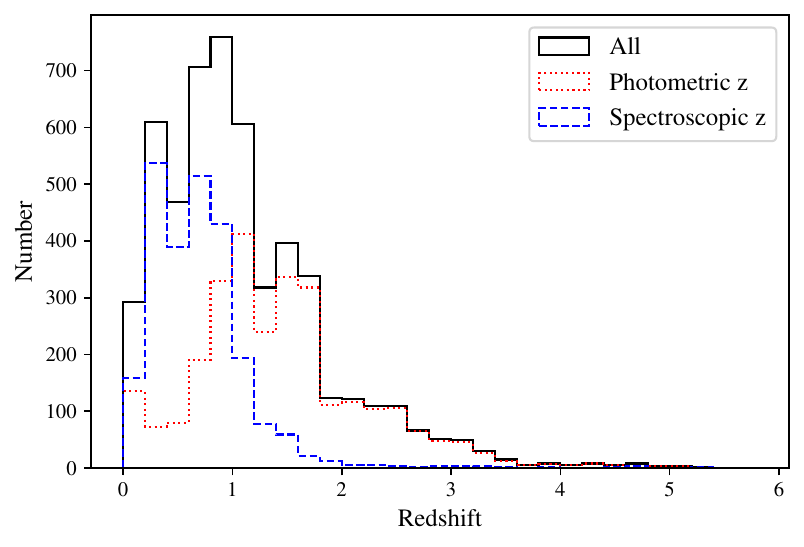}
\caption{Redshift distribution of the $5\,223$ objects in cross-matched sample. The distribution of objects with spectroscopic redshifts ($2\,427$ objects) can be seen as a blue dashed line, whereas the distribution of those with photometric redshifts ($2\,796$ objects) can be seen as the red dotted line.}
\label{HISTOZ}
\end{figure}

\begin{figure}
\includegraphics[width=\columnwidth]{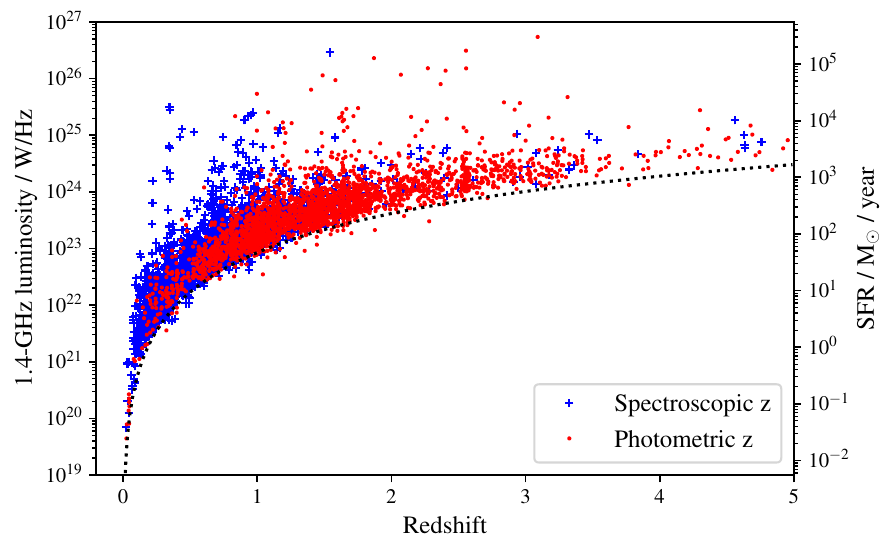}
\caption{The rest-frame $1.4$ GHz radio luminosity - redshift distribution of the cross-matched sample. Objects with spectroscopic redshifts can be seen as blue crosses and those with photometric redshifts as red dots. The dotted black line represents a flux limit of $20~\muup$Jy. The y axis on the right hand side shows an estimate of the star-formation rate, scaled from the radio luminosity using the \citet{Bell2003} relation.}
\label{LUMZPLOT}
\end{figure}

The sample presented in this paper contains 5223 visually cross-matched sources, which is 86 per cent of the parent radio sample. Spectroscopic redshifts are available for 2427 sources, and for the remaining 2796 sources we use photometric redshifts from \citet{Hatfield2022} (see Section~\ref{section:multi-data} for details). 
Fig.~\ref{HISTOZ} displays the redshift distribution of our cross-matched sample for objects with spectroscopic redshifts (blue dashed line) and photometric redshifts (red dotted line). The median redshifts for the spectroscopic, photometric and entire sample are $z = 0.66$, $z = 1.36$ and $z = 0.94$ respectively. To calculate rest-frame $1.4$ GHz radio luminosities of our radio sample, we assume a spectral index of $\alpha = 0.7$ (where $S \propto \nu^{-\alpha}$). Due to the wide bandwidth of the MeerKAT L-band receivers and the varying response of the primary beam with frequency the effective frequency of the MIGHTEE data varies across the image. This is discussed in detail in \citet{Heywood2022}, and we use the effective frequency map released with that work to scale the MIGHTEE flux densities and luminosities to 1.4~GHz. The luminosity-redshift plot of the objects in our sample is shown in Fig.~\ref{LUMZPLOT}.  This shows that we are able to investigate the evolution of faint ($L_{1.4} \sim 10^{24}$\,W\,Hz$^{-1}$) AGN out to the epoch of re-ionisation and assuming the correlation between SFR and radio luminosity \citep[e.g.][]{YunReddyCondon2001,Bell2003,Delvecchio2021} star-forming (SFR $\sim 50$\,M$_{\odot}$\,yr$^{-1}$) and starburst (SFR $> 100$\,M$_{\odot}$\,yr$^{-1}$) galaxies to $z\sim 1$ and $z \sim 5$, respectively, if the optical and near-infrared data are deep enough to measure redshifts.

It tends to be more difficult to produce accurate photometric redshift estimates for radio sources, due to the prevalence of bright emission lines in both star-forming galaxies and AGN, and the possible AGN contribution to the continuum. We therefore assess the accuracy of the photometric redshifts of our sample by comparing sources that have both spectroscopic ($z_\textrm{Spec}$) and photometric redshifts ($z_\textrm{Photo}$) available. 

The spread between the two redshift estimates can be defined as $\Delta z/(1+z_\textrm{Spec})$ where $\Delta z = z_\textrm{Spec}-z_\textrm{Photo}$. As in \cite{Ilbert2006} and \cite{Jarvis2013}, we calculate the normalized median absolute deviation (NMAD) as ${\rm NMAD} = 0.023$ which implies there is a good agreement between the two quantities.  
Defining outliers as cross-matches that have $|z_\textrm{Spec}-z_\textrm{Photo}|/(1+z_\textrm{Spec}) > 0.15$, we find that only $115$ objects or $4.94$ per cent of the sample have poorly determined photometric redshifts, showing that the photometric redshifts are fairly robust. In the future, spectroscopic redshifts for further MIGHTEE sources will become available from the Deep Extragalactic VIsible Legacy Survey \citep[DEVILS,][]{Davies2018}, the Multi-Object Optical and Near-infrared Spectrograph \citep[MOONS,][]{ Cirasuolo2012} and the 4-metre Multi-Object Spectroscopic Telescope \citep[4MOST,][]{DeJong2019},  and in particular the Optical, Radio Continuum and HI Deep Spectroscopic Survey \citep[ORCHIDSS; ][]{ORCHIDSS}.

\section{Comparisons with simulations}
\label{section:sims}

In this section we compare the radio flux densities and redshift distributions of the AGN and star-forming galaxies (SFG) in our visually cross-matched sample to those from the Square Kilometre Array Design Study (SKADS) \citep{Wilman2008, Wilman2010} and the more recent Tiered Radio Extragalactic Simulation \citep[T-RECS;][]{Bonaldi2019}. We use the AGN and star-forming galaxy classifications from \cite{Whittam2022}, which make use of the abundance of multi-wavelength data available in the COSMOS field to classify sources as AGN and SFG. As these classification are only available for the visually cross-matched sample, we restrict our analysis to that sample for the remainder of this section. The classification scheme is described in detail in \citet{Whittam2022} and outlined briefly here. The classifications are based on five criteria which are then combined to give an overall classification. The first diagnostic makes use of the far-infrared-radio correlation to identify objects with significantly more radio emission than would be expected from star-formation alone. Following \cite{Delvecchio2021}, sources with radio emission $>2\sigma$ above the correlation are classified as AGN. The second diagnostic identifies AGN from their X-ray emission. Objects with X-ray luminosities of $L_{X} > 10^{42}$ erg s$^{-1}$ are classified as AGN. Third, AGN are identified from their mid-infrared colours using a colour-colour diagram as described in \cite{Donley2012}. For the fourth diagnostic, sources detected by Very Long Baseline Interferometry (VLBI) observations of the COSMOS field by \cite{HerreraRuiz2017} are labelled as AGN. Finally, objects that have point-like morphologies at optical wavelengths (using Hubble ACS I-band data) are classified as AGN. A source is classified as an AGN if it meets any one (or more) of the five AGN criteria. Sources which we can securely classify as not being an AGN using all five criteria are classified as star-forming galaxies. The depth of the X-ray data used means that we can only rule out AGN-related X-ray emission at redshifts $< 0.5$, meaning that we are only able to securely classify objects as star-forming galaxies in this redshift range. We therefore introduce an additional classification of `probable SFG' for sources which have redshifts $> 0.5$ so are unable to fulfil the `not X-ray AGN' criteria, but which are classified as `not AGN' using the other four criteria. For the remainder of this work we combine the SFG and `prob SFG' classes and refer to the combination simply as `SFG'. The AGN are further classified as radio-loud and radio-quiet. All AGN which meet the `radio excess' criteria are considered to be radio-loud, while those which do not have excess radio emission, but are classified as an AGN using one of the other criteria are classified as radio-quiet AGN.

\begin{figure}
\includegraphics[width=\columnwidth]{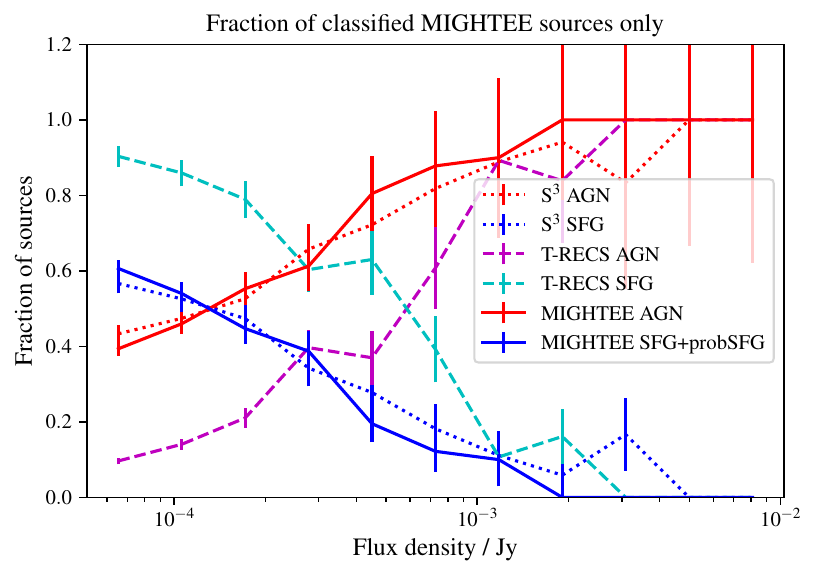}
\includegraphics[width=\columnwidth]{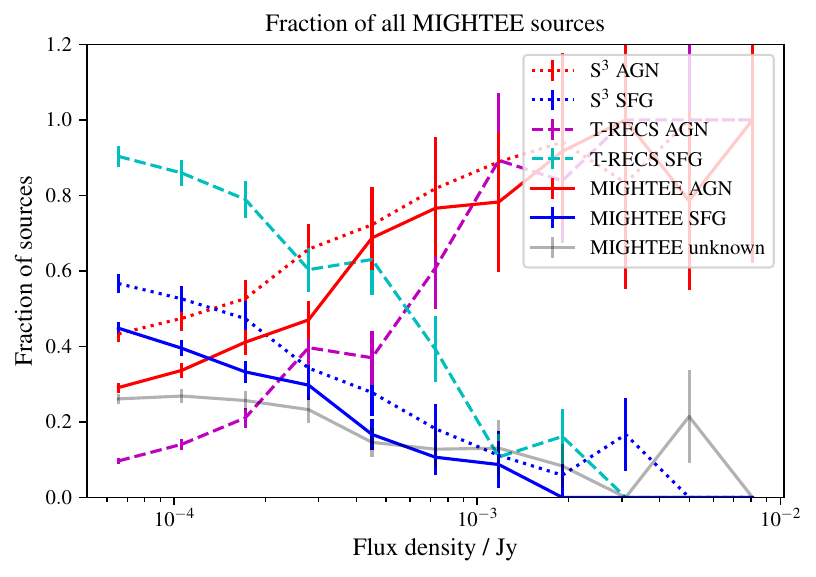}
\includegraphics[width=\columnwidth]{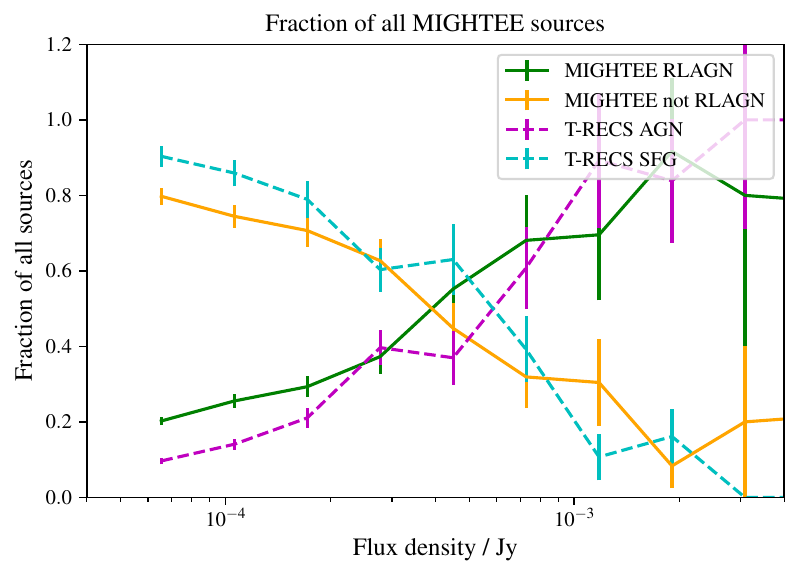}
\caption{Fraction of AGN and SFG as a function of 1.4-GHz flux density in the visually cross-matched MIGHTEE sample compared to the SKADS simulated skies and T-RECS. The top panel shows the number of AGN and SFG (including `probable SFG' for MIGHTEE) as a fraction of the classified radio sources - i.e.\ only MIGHTEE sources which we are able to classify as either AGN or SFG are included.  The middle panel shows the MIGHTEE SFG and AGN as a fraction of all MIGHTEE sources (including unmatched and unclassified sources). The unclassified and unmatched MIGHTEE sources are shown as the pale grey line (labelled `MIGHTEE unknown'). The bottom panel shows the fraction of MIGHTEE radio-loud AGN and all sources not classified as radio-loud AGN, compared to the AGN and SFG in T-RECS. Note that the `MIGHTEE not AGN' class includes all unmatched and unclassified sources, as well as those sources classified as SFG and radio-quiet AGN. MIGHTEE fluxes are scaled to 1.4 GHz assuming a spectral index of 0.7. Uncertainties shown are Poisson errors.}
\label{fig:fraction-S3}
\end{figure}

\begin{figure}
\includegraphics[width=\columnwidth]{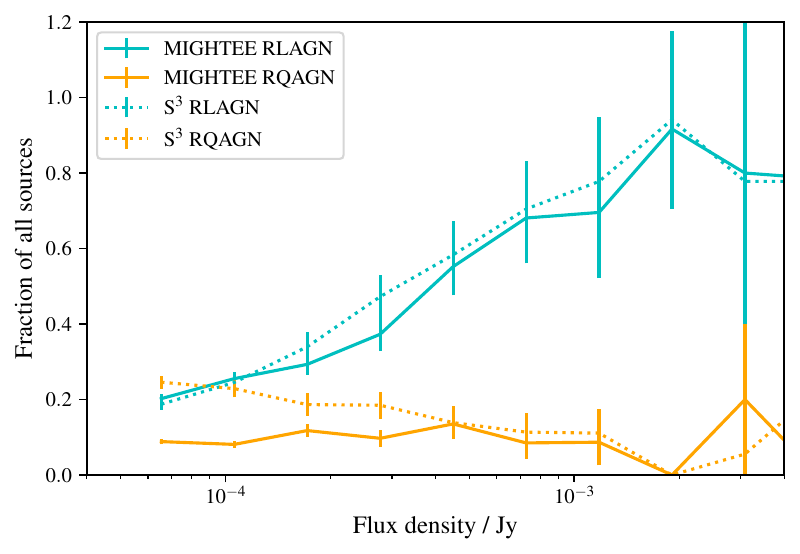}
\caption{The fraction of sources classified as radio-loud and radio-quiet AGN in MIGHTEE and SKADS.  MIGHTEE fluxes are scaled to 1.4 GHz assuming a spectral index of 0.7. Uncertainties shown are poisson errors.}
\label{fig:fraction-S3-RL}
\end{figure}

\subsection{Flux distribution}\label{section:sims-flux}

Fig.~\ref{fig:fraction-S3} shows the fraction of AGN and star-forming galaxies as a function of their total radio flux density, compared to the SKADS and T-RECS simulations.
The MIGHTEE flux densities have been scaled to $1.4$ GHz using the effective frequency map assuming a spectral index of $\alpha = 0.7$. So as not to be affected by incompleteness due to the variation in noise across the MIGHTEE image, we cut all three catalogues at $S_{1.4~\textrm{GHz}} = 50~\muup$Jy as the MIGHTEE sample is complete above this flux density (see \citealt{Hale2022}). With this flux density cut applied, the MIGHTEE sample contains 3294 sources, of which 2824 (86 per cent) have a multi-wavelength counterpart identified in the visually cross-matched catalogue. 2467 (75 per cent) of these objects are classified as an AGN or SFG, the remaining sources do not have enough multi-wavelength information available to be able to securely classify them.

The top panel of Fig.~\ref{fig:fraction-S3} shows the fraction of classified MIGHTEE sources which are identified as AGN or SFG as a function of 1.4-GHz flux density. This demonstrates that AGN and star-forming fractions in the MIGHTEE sample are in good agreement with the SKADS simulations. Both show that the AGN fraction increases with increasing radio flux density from $\sim~40$ per cent at $\sim50\,\muup$Jy to $\sim 95$ per cent at $1$~mJy. Both SKADS and our sample show equal fractions for SFG and AGN at $\sim100 \muup$Jy, below which SFG become the dominant population. This is consistent with the findings of \cite{Padovani2015} who find that SFG become the dominant population below $\sim 100 \muup$Jy using radio observations of the Extended Chandra Deep Field South (E-CDFS) Very Large Array sample, as well as \cite{Smolcic2017b}, using $3$ GHz observations of the COSMOS field. 

In contrast, T-RECS significantly over-predicts the fraction of SFGs, and therefore under-predicts the fraction of AGN, when compared to the MIGHTEE sample. However, this plot does not include the sources we are unable to classify; both those with an optical match but without enough information to classify as AGN or SFG, and those without an optical match. The middle panel of Fig.~\ref{fig:fraction-S3} show the proportion of MIGHTEE SFG and AGN in the full MIGHTEE sample, with the fraction of sources without a classification shown by the grey line. This shows that even if none of the unclassified MIGHTEE sources are AGN, the fraction of AGN at radio flux densities less than 1~mJy is higher in the MIGHTEE sample than predicted by the T-RECS simulation. At $S_{1.4_\textrm{GHz}} \sim 50~\muup$Jy around 30 per cent of the MIGHTEE sample are AGN (and this should be considered a lower limit on the fraction of AGN, as it is very possible that some of the unknown sources are AGN), while T-RECS predicts that less than 10 per cent of this sample should be AGN. Note that despite their faint radio flux densities the majority of these AGN are not radio quiet - they have an excess over what would be expected from star-formation alone. This can been seen in Fig.~\ref{fig:fraction-S3-RL}.  

However, the T-RECS work does not include radio-quiet AGN (which are instead included in the SFG class) which could account for some of this difference. To test this, in the bottom panel of Fig.~\ref{fig:fraction-S3} we show the MIGHTEE radio-loud AGN (in yellow) and all other MIGHTEE sources not classified as radio-loud AGN (green line, this includes radio-quiet AGN, SFG and unclassified sources). This shows that even when radio-quiet AGN are included and all unclassified sources are assumed to be SFG, T-RECS still significantly over-predicts the fraction of SFG in the observed sample by $\sim 10$\,per cent at $S_{1.4}  \lesssim 0.5$\,mJy.

In Fig.~\ref{fig:fraction-S3-RL} we show the fraction of radio-loud and radio-quiet AGN in the MIGHTEE sample compared to what is predicted by SKADS as a function of flux density. There is a reasonable agreement between the two samples, although the MIGHTEE sample contains fewer radio-quiet AGN than predicted at $S_{1.4~\textrm{GHz}} < 200 \muup$Jy. There have been several studies investigating the process responsible for radio emission in radio-quiet AGN. For example, \cite{Kimball2011} and \cite{Kellerman2016} find that using radio observations of radio-quiet quasars that a significant fraction of the radio emission could be attributed to star formation. On the other hand, \cite{White2015,White2017} use multi-wavelength data to fit the spectral energy distribution of a sample of radio-quiet AGN from blank surveys and targeted surveys to determine the contribution of star formation to the radio luminosity and find that the AGN is responsible for the bulk of the radio emission. More recent work \citep[e.g.][]{Macfarlane2021,Xiao2022} also attribute the bulk of the radio emission in radio-quiet quasars as due to similar jet-production processes occurring in their radio-loud counterparts. Therefore, it is clear that more work is needed in this area, and classifying such faint radio sources as AGN requires very good ancillary data. For example, past work has been concentrated on the specific class of radio-quiet quasars, where the nuclear point source at optical wavelengths is dominant, whereas the classifications here include mid-infrared and X-ray data.

\begin{figure}
\includegraphics[width=\columnwidth]{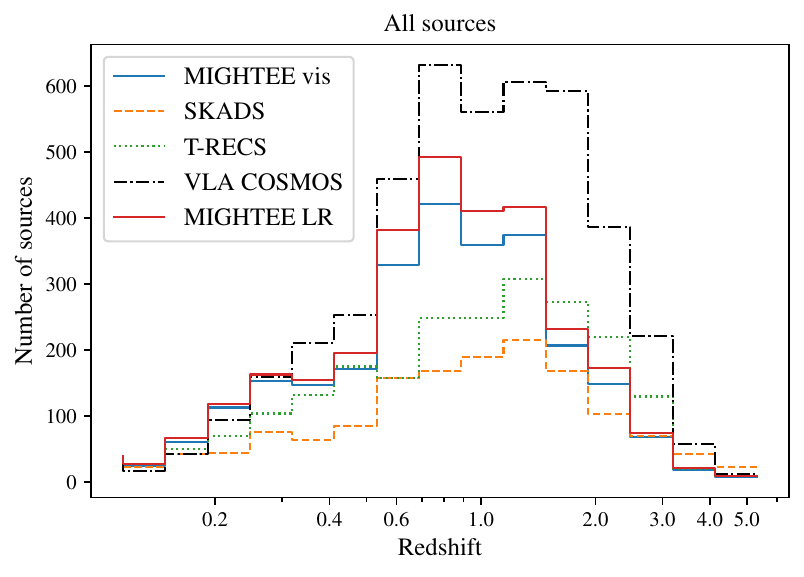}
\includegraphics[width=\columnwidth]{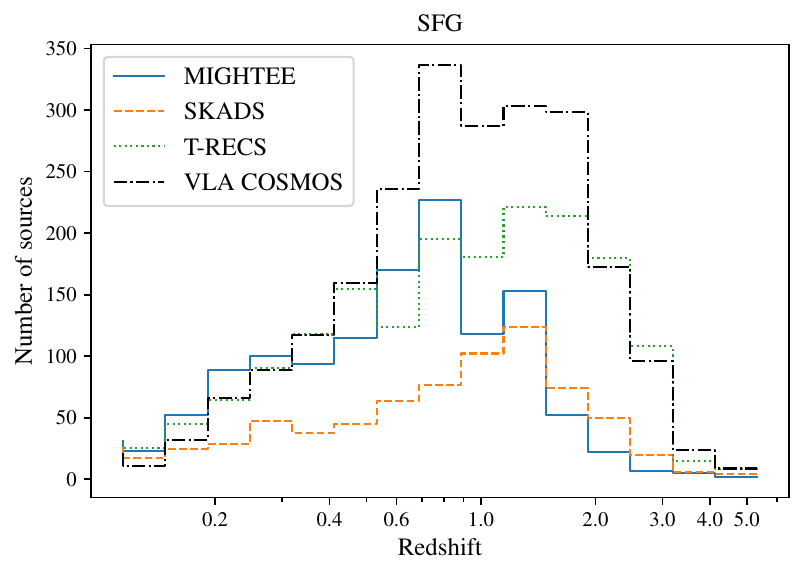}
\includegraphics[width=\columnwidth]{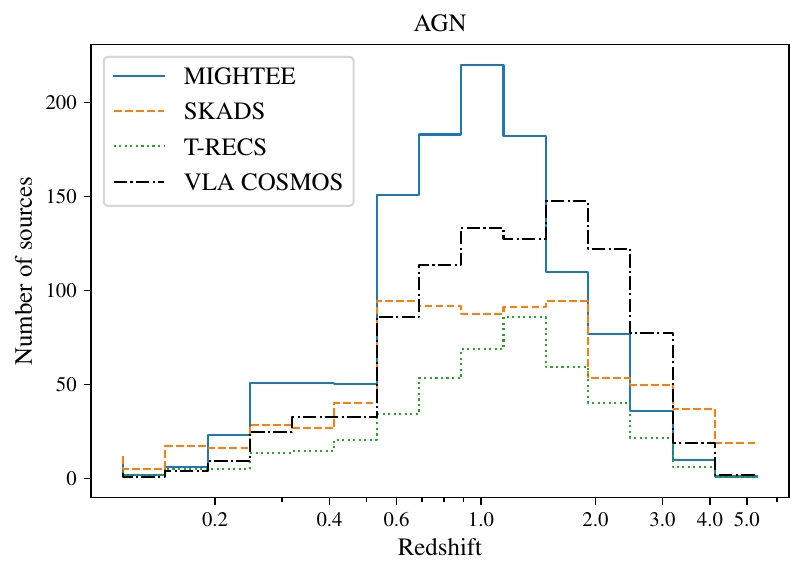}
\caption{Comparison between the redshift distribution of simulated radio sources from SKADS (dashed lines), T-RECS (dotted lines) and the MIGHTEE visually cross-matched sample (solid lines). The \citet{Smolcic2017b} VLA-COSMOS 3 GHz sample is also shown (dot-dashed line). All distributions are normalised to the MIGHTEE area of 0.86~deg$^2$. The top panel shows all sources, the middle panels shows sources classified as SFG, and the bottom panel shows AGN. See text for details of classifications. The distribution of sources in the LR-matched MIGHTEE catalogue, which contains more sources, is also shown by the red solid line in the top panel.}
\label{fig:simsz}
\end{figure}

\begin{figure}
    \centering
    \includegraphics[width=\columnwidth]{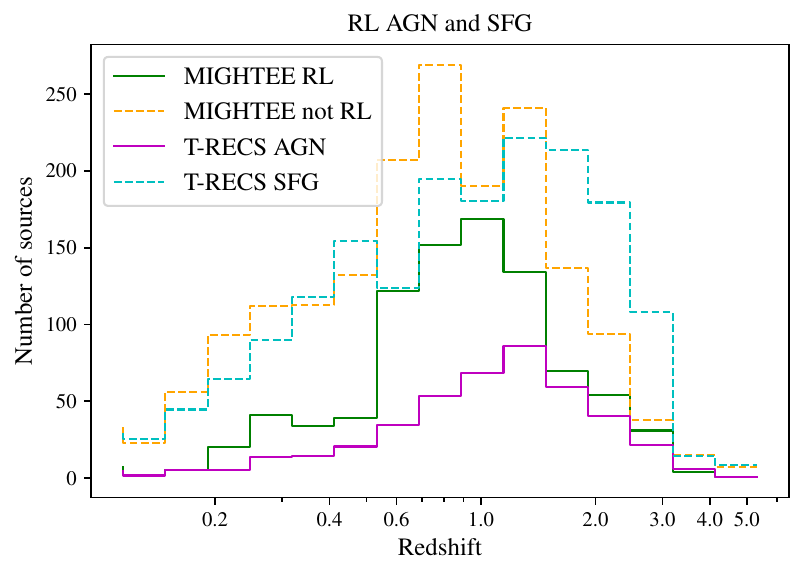}
    \caption{Comparison of the redshift distribution of simulated radio sources from T-RECS and the MIGHTEE visually cross-matched sample. For the T-RECS sample, AGN and SFG are show separately (T-RECS does not include radio-quiet AGN, so these will be included in the `SFG' class). For the MIGHTEE sample, radio-loud AGN and shown in green, and all other sources (i.e.\ radio-quiet AGN, SFG and unclassified sources) are shown in magenta. }
    \label{fig:simsz_RLAGN}
\end{figure}

\subsection{Redshift distribution}

Fig.~\ref{fig:simsz} shows the redshift distributions for the MIGHTEE AGN and star-forming galaxies in the visually cross-matched sample, compared with those from the SKADS and T-RECS simulations. We use the spectroscopic redshifts where available, and the photometric redshifts for all other sources. The distributions from the \citeauthor{Smolcic2017b} VLA-COSMOS 3 GHz work are also shown for reference\footnote{The VLA-COSMOS team define a SFG class, which includes some sources with a radio excess, and a `clean SFG' class where these radio-excess sources are removed. Here we compare to their `clean SFG' class, as this is more consistent with the definition of SFG used in this work, which excludes all radio-excess sources.}. As above, the MIGHTEE SFG class shown here is a combination of the `SFG' and `probable SFG' classes described in \citet{Whittam2022}.  All distributions are normalised to the MIGHTEE area used in this work, 0.86 deg$^2$. The visually cross-matched MIGHTEE sample in the top panel only includes the 2824 sources that we are able to identify a host for (86 per cent of the full sample with $S_{1.4~\textrm{GHz}} > 50~\muup$Jy). As we are able to cross-match a higher proportion of sources using the Likelihood Ratio method (3088 sources, 94 per cent of the full sample with $S_{1.4~\textrm{GHz}} > 50~\muup$Jy), we also show the redshift distribution of the LR-matched sample for reference in the top panel of Fig.~\ref{fig:simsz}. While they are more complete, these identifications are not as robust as those from the visually cross-matched catalogue. However, their distribution gives a good indication of the potential distribution of the sources missing from the visually cross-matched sample so provides a useful reference. We note that the AGN and SFG classifications are not currently available for the LR matched catalogue, so we are not able to include this catalogue on the bottom two panels of Fig.~\ref{fig:simsz}.

The differences between our observed distribution of sources and the simulated distributions are highlighted when the source populations are split into SFG and AGN. There are more AGN in the MIGHTEE sample at $z \sim 1$ than predicted by either simulation. T-RECS under-predicts the number of AGN to a greater extent than SKADS, this is probably because the T-RECS `AGN' class only includes radio-loud AGN, as discussed in Section~\ref{section:sims-flux}. To account for this, in Fig.~\ref{fig:simsz_RLAGN} we show the T-RECS AGN and SFG compared to the MIGHTEE radio-loud AGN and all other MIGHTEE sources (i.e.\ radio-quiet AGN, SFG and unclassified sources), which should be more directly comparable classifications. This shows that T-RECS still under-predicts the number of AGN at $z \sim 1$, even when only radio-loud AGN are considered.

We note that as the MIGHTEE sample shown in Fig.~\ref{fig:simsz} only includes sources we are able to securely classify, the number of AGN (and SFG) shown here should be considered a lower limit. The  prescriptions for simulating the AGN population in both simulations are based on observations at higher fluxes and extrapolated down to the fluxes reached by the MIGHTEE survey. For example, SKADS uses the Fanaroff and Riley type I and II (FRI and FRII; \citealt{Fanaroff1974}) evolution models from \citet{Willott2001}, along with the observed relationship between X-ray and radio luminosity for radio-quiet quasars \citep{Brinkmann2000}, then extrapolate these to fainter flux densities. This work demonstrates that there are more AGN than predicted by these extrapolations. These AGN that are missing from the simulations are predominately low-excitation radio-loud AGN \citep[LERGs, see e.g. ][]{Heckman2014}, which show an excess of radio emission but do not display the other indicators of AGN emission typically present in more highly-accreting nuclei, such as strong nuclear emission and mid-IR emission from a dusty torus. It is only due to the combination of deep radio data and excellent multi-wavelength data in the MIGHTEE fields that we are able to identify these very faint AGN. This has implications for the role of radio galaxies in galaxy evolution, as it suggests that mechanical feedback could play a significant role even at faint flux densities. This is discussed further in \citet{Whittam2022}.

In terms of SFG, both the MIGHTEE and VLA-COSMOS observed samples show good agreement with T-RECS at $z < 1$. The SKADS simulation, however, under-predicts the number of SFG observed at $z \lesssim 0.6$. This is in agreement with the growing evidence in the literature that the SKADS simulation underestimates the number of SFG at faint flux densities ($S_{1.4~\textrm{GHz}} \lesssim 0.1$~mJy), see e.g.\ \citealt{Smolcic2017a,prandoni2018,Mauch2020,Matthews2021,Hale2022}. While the absolute numbers of SFG and AGN in SKADS do not agree well with the observations, as discussed in this section, the fractions of AGN and SFG in SKADS are in good agreement with the observed fractions, as shown in Fig.~\ref{fig:fraction-S3} and discussed in Section~\ref{section:sims-flux}. This is because SKADS does not include a significant number of faint radio AGN at similar redshifts to the SFGs. As the MIGHTEE observations only cover a relatively small area (0.86~deg$^2$), it is possible the cosmic variance has an impact on the absolute number of sources in the field. However, \citet{Heywood2013} shows that cosmic variance is only expected to around $\sim5$ per cent of the number density at $S_{1.4~\textrm{GHz}} \sim 100\muup$Jy. Additionally, \citet{Hale2022} demonstrates that the MIGHTEE source counts in the COSMOS field (used in this work) are consistent with those from the XMM-LSS field.

On the other hand, the VLA COSMOS sample appears to contain more SFGs and fewer AGN at $z \sim 1$ than the MIGHTEE sample. This is probably primarily due to differences in the methods used to classify the sources, particularly the different criteria used to identify radio-excess AGN. When comparing the sources in common, there are a number of radio-excess AGN in the MIGHTEE sample which are identified as SFG in the VLA COSMOS work. This is in part because the VLA COSMOS team require a source to have a $3\sigma$ radio excess to be classified as radio loud \citep{Smolcic2017b}, while we follow the more recent work by \citet{Delvecchio2021} and only require a $2\sigma$ radio excess. This results in a higher completeness, but could cause a $3 - 4$ per cent contamination of SFG in the radio-loud AGN sample. This is discussed in detail in \citet{Whittam2022} where the classification schemes for the two studies are compared. In the lowest redshift bins ($z < 0.3$) MIGHTEE detects more SFG (and more sources in total) than VLA COSMOS. As discussed in \citet{Hale2022} there are a number of extended galaxies detected in MIGHTEE which are not detected in VLA-COSMOS despite having total flux densities above their detection limit. This is because the configuration of the VLA used for the VLA-COSMOS observations lacks short baselines, so while it provides excellent resolution, it is not sensitive to extended emission, resulting in these sources being missed.

\section{Conclusions}
\label{Conclusions}

In this work\footnote{This paper is referred to as Prescott et al. (in prep) in \citet{Whittam2022} and \citet{Hale2022}.} we have cross-matched the MIGHTEE Early Science radio catalogue in the central part COSMOS field with a multi-wavelength catalogue of objects selected in the near-infrared $K_s$-band both by eye and by using an automated Likelihood Ratio method. The cross-matched catalogues are released with this work. Our main results can be summarised as follows:

\begin{itemize}
\item From an initial {\sc pybdsf} catalogue of $6\,102$ radio sources, we find that $5\,223$ radio sources can be successfully assigned to a multi-wavelength counterpart via visual inspection.

\item We compare our visually cross-matched sample to samples obtained using the likelihood ratio method. With the automated LR method we are able to identify counterparts for 94 per cent of radio components in the low-resolution MIGHTEE catalogue, and these matches agree with those identified visually in 95 per cent of cases. The fraction we are able to match rises to 97 per cent when we consider sources which are unresolved only.

\item Visual inspection is still crucial for cross-matching extended and multi-component radio sources, and for identifying confused sources. The LR method is only able to match 61 per cent of sources with $S_{1.4~\textrm{GHz}} > 100~\muup$Jy, while using visual inspection we are able to identify counterparts for 93 per cent of the same sample. This highlights the benefits of combining the two methods; by using the LR we are able to automatically match a large number of the fainter, compact sources, but visual inspection is necessary to match the extended, complex sources. A dual approach of automated and visual inspection will be implemented for future MIGHTEE observations of the remainder of the COSMOS field and the XMM-LSS, E-CDFS and ELAIS-S1 fields. 

\item Our sample contains a mixture of AGN and star-forming galaxies, which can be probed out to $z \sim 5$. We show that the fractions of AGN and star-forming galaxies as function of radio flux agree well with SKADS simulations, with star-forming galaxies becoming the dominant population below flux densities of $\sim100\muup$Jy. The T-RECS simulation, however, seems to under-predict the fraction of AGN and over-predicts the fraction of SFG below $S_{1.4~\textrm{GHz}} \sim 1$~mJy.

\item The MIGHTEE sample contains more AGN at $z \sim 1$ than predicted by either simulation (although SKADS is closer to matching the observed distribution than T-RECS). The majority of these AGN are low-excitation radio galaxies (LERGs) and it is only due to the combination of deep radio data and excellent multi-wavelength data in the MIGHTEE field that we are able to detect these faint AGN.

\end{itemize}

\section*{Acknowledgements}

We thank the referee for their helpful comments which greatly improved this manuscript. The MeerKAT telescope is operated by the South African Radio Astronomy Observatory, which is a facility of the National Research Foundation, an agency of the Department of Science and Innovation.
IH, MJJ, PWH and IHW acknowledges generous support from the Hintze Family Charitable Foundation through the Oxford Hintze Centre for Astrophysical Surveys.
MP acknowledges support by the South African Square Kilometre Array Project, the South African National Research Foundation and the Inter-University Institute for Data Intensive Astronomy (IDIA). IDIA is a partnership of the University of Cape Town, the University of Pretoria, the University of the Western Cape and the South African Radio Astronomy Observatory. 
CLH acknowledges support from the Leverhulme Trust through an Early Career Research Fellowship.
MJJ acknowledges support of the STFC consolidated grant [ST/S000488/1] and [ST/W000903/1] and from a UKRI Frontiers Research Grant [EP/X026639/1]. 
MG was partially supported by the Australian Government through the Australian Research Council's Discovery Projects funding scheme (DP210102103).
NM acknowledges the support of the LMU Faculty of Physics.
LM and MV acknowledge support from the Italian Ministry of Foreign Affairs and International Cooperation (MAECI Grant Number ZA18GR02) and the South African Department of Science and Innovation's National Research Foundation (DSI-NRF Grant Number 113121) under the ISARP RADIOSKY2020 Joint Research Scheme.
MGS acknowledges support from the South African Radio Astronomy Observatory and National Research Foundation (Grant No. 84156).
This work was supported by the Medical Research Council [MR/T042842/1]. 
This research made use of {\sc aplpy}, an open-source plotting package for Python (Robitaille and Bressert, 2012), hosted at \url{http://aplpy.github.com}. We also acknowledge the IDL Astronomy User's Library, and IDL code maintained by D.~Schlegel (IDLUTILS) as valuable resources.

\section*{Data Availability}

The release of the MIGHTEE Early Science continuum data used for this work is discussed in depth in \citet{Heywood2022}, details of the data release and how to access the data are provided there. The catalogue presented in this article is summarised in Appendix~\ref{appendixa} and is published online with this article as online-only material.
The AGN / SFG classification catalogue was released with \cite{Whittam2022}.

\bibliographystyle{mnras}
\bibliography{BIB}
\hspace{1cm}

\noindent {\it Author Affiliations}

% List of institutions
\noindent $^{1}$ Astrophysics, Department of Physics, University of Oxford, Keble Road, Oxford, OX1 3RH, UK 
\\
$^{2}$ Department of Physics and Astronomy, University of the Western Cape, Robert Sobukwe Road, 7535 Bellville, Cape Town, South Africa. 
\\
$^{3}$ Inter-University Institute for Data Intensive Astronomy (IDIA), University of the Western Cape, Robert Sobukwe Road, 7535 Bellville, Cape Town, South Africa.
\\
$^{4}$ Institute for Astronomy, Royal Observatory Edinburgh, Blackford Hill, Edinburgh, EH9 3HJ, UK
\\
$^{5}$ Centre for Radio Astronomy Techniques and Technologies, Department of Physics and Electronics, Rhodes University, PO Box 94, Makhanda, 6140, South Africa. 
\\
$^{6}$ South African Radio Astronomy Observatory, 2 Fir Street, Black River Park, Observatory, Cape Town, 7925, South Africa.
\\
$^{7}$ The Inter-University Institute for Data Intensive Astronomy (IDIA), Department of Astronomy, University of Cape Town, Private Bag X3, Rondebosch, 7701, South Africa.
\\
$^{8}$ International Centre for Radio Astronomy Research, Curtin University, Bentley, WA 6102, Australia
\\
$^{9}$ University Observatory, Faculty of Physics, Ludwig-Maximilians-Universit\"at, Scheinerstr. 1, 81679 Munich, Germany
\\
$^{10}$ Department of Astronomy, University of Cape Town, 7701 Rondebosch, Cape Town, South Africa
\\
$^{11}$ INAF - Istituto di Radioastronomia, via Gobetti 101, 40129 Bologna, Italy
\\
$^{12}$ Centre for Extragalactic Astronomy, Department of Physics, Durham University, Durham, DH1 3LE, UK
\\
$^{13}$ Institute for Computational Cosmology, Department of Physics, Durham University, Durham, DH1 3LE, UK
\\
$^{14}$ Jodrell Bank Centre for Astrophysics, University of Manchester, Oxford Road, Manchester, UK
\\
$^{15}$ Instituto de Astrof\'isica e Ci\^encias do Espa\c{c}o, Universidade de Lisboa, OAL, Tapada da Ajuda, PT1349-018 Lisboa, Portugal
\\
$^{16}$ Departamento de F\'isica, Faculdade de Ci\^encias, Universidade de Lisboa, Edif\'icio C8, Campo Grande, Portugal
\\
$^{17}$ Purple Mountain Observatory and Key Laboratory for Radio Astronomy, Chinese Academy of Sciences, Nanjing, China
\\
$^{18}$ School of Astronomy and Space Science, University of Science and Technology of China, Hefei, China
\\
$^{19}$ Physics Department, University of Johannesburg, 5 Kingsway Ave, Rossmore, Johannesburg, 2092, South Africa
\\
$^{20}$ South African Astronomical Observatory, P.O. Box 9, Observatory 7935, Cape Town, South Africa
\\
$^{21}$ A\&A, Department of Physics, Faculty of Sciences, University of Antananarivo, B.P. 906, Antananarivo 101, Madagascar
\\
$^{22}$ Centre for Astrophysics Research, University of Hertfordshire, College Lane, Hatfield, AL10 9AB, UK\\

\appendix

\section{Structure of the Cross-Matched Catalogue}
\label{appendixa}

The Level-1 MIGHTEE Early Science catalogue of radio sources was released in \cite{Heywood2022}. The cross-matched catalogues used in this work are released here, and are known as the Level-2 catalogues. We release three catalogues; the visually cross-matched catalogue described in Section~\ref{Cross-Matching} (based on the low-resolution Level-0 catalogue), and two catalogues cross-matched using the Likelihood Ratio method described in Section~\ref{section:LR}. One LR catalogue is based on the low-resolution Level-0 catalogue, and one is based on the high-resolution catalogue. The Level-3 catalogue contains source type classifications and was released with \citet{Whittam2022}. The three catalogues released with this work follow the same structure, which is described below.

\noindent \textbf{(0)}: Name: An IAU-style identifier of the form JHHMMSS.SS+/- DDMMSS.S, based on the position of the host galaxy. 

\noindent \textbf{(1)}: RA\_Radio: The J2000 Right Ascension of the radio source in degrees from the {\sc pybdsf} catalogue. If this is multiple component radio source this is the R.A. of the brightest component.  

\noindent \textbf{(2)}: DEC\_Radio: The J2000 Declination of the radio source in degrees from the {\sc pybdsf} catalogue. If this is multiple component source this is the Declination of the brightest component.  

\noindent \textbf{(3)}: RA\_Host: The J2000 Right Ascension of the $K{_s}$-band selected counterpart from the multi-wavelength catalogue. 

\noindent \textbf{(4)}: DEC\_Host: The J2000 Declination of the $K{_s}$-band selected counterpart in degrees from the multi-wavelength catalogue.

\noindent \textbf{(5)}: Peak\_Flux: The peak radio flux of the source from the {\sc pybdsf} catalogue.   

\noindent \textbf{(6)}: Peak\_Flux\_err: Error on the peak radio flux of the source from the {\sc pybdsf} catalogue. 

\noindent \textbf{(7)}: Total\_Flux: The total radio flux density of the source from the {\sc pybdsf} catalogue.

\noindent \textbf{(8)}: Total\_Flux\_err: The total radio flux density and error of the source from the {\sc pybdsf} catalogue.

\noindent \textbf{(9)}: Phot\_z: photometric redshift of the host galaxy.

\noindent \textbf{(10)}: Phot\_z\_err: error on the photometric redshift of host galaxy.

\noindent \textbf{(11)}: Spec\_z: Spectroscopic redshift of the source if available. 

\noindent \textbf{(12)}: Spec\_z\_err: error of the radio source if available.

\noindent \textbf{(13)}: Spec\_z\_note: Source of the spectroscopic redshift.

\smallskip
\noindent Visually cross-matched catalogue only:

\noindent \textbf{(14)}: N\_Comp: The number of {\sc pybdsf} radio components that the object comprises of. 

\noindent \textbf{(15)}: FLAG: Additional information about the match. 100 = single component, 120 = multi-component source, 112 = split source. 

\smallskip
\noindent LR catalogues only:

\noindent \textbf{(14)}: LR: The likelihood ratio for that match. 

\noindent \textbf{(15)}: GoodLR: indicates whether or not to accept the LR match. =1 if LR $> L_\textrm{thr}$, = 0 otherwise.

% Don't change these lines
\bsp	% typesetting comment
\label{lastpage}
\end{document}